\documentclass[a4paper,11pt]{article}
\usepackage{jheppub}
\usepackage[utf8x]{inputenc}
\usepackage[T1]{fontenc}
\usepackage{braket}
\usepackage{physics}
\usepackage{color}
\usepackage{amsfonts}
\usepackage{amssymb}
\usepackage{amsmath}
\usepackage[normalem]{ulem}
\usepackage[font={footnotesize}]{caption}
\usepackage{graphicx}
\usepackage{epstopdf}
\usepackage{enumitem}
\usepackage{subcaption}
\usepackage{comment}
\usepackage{mathtools}
\usepackage{ragged2e}
\usepackage{hyperref}
\hypersetup{
	colorlinks=true,
	linkcolor=blue,
	filecolor=cyan, 
	urlcolor=blue,
	citecolor=red,
} 
\usepackage[titletoc,toc,title]{appendix}
\numberwithin{equation}{section}
\usepackage[nameinlink]{cleveref}
\usepackage{float}
\usepackage{makecell,booktabs}
\usepackage{color, colortbl}
\definecolor{Gray}{gray}{0.9}
\usepackage{multirow}
\makeatletter
\newcommand{\vast}{\bBigg@{3}}
\newcommand{\Vast}{\bBigg@{5}}
\makeatother

\newcommand{\TTbar}{\text{T}\bar{\text{T}}}

\newcommand{\zbar}{\raisebox{0.2ex}{--}\kern-0.6em Z}

\def\CC{{\cal C}}
\def\CD{{\cal D}}
\def\CE{{\cal E}}
\def\CF{{\cal F}}

\def\CK{{\cal K}}
\def\CL{{\cal L}}

\def\CM{{\cal M}}

\def\CO{{\cal O}}

\def\CS{{\cal S}}

\def\d{\textrm{d}}
\def\del{\partial}

\def\TTbar{\textrm{T}\overline{\textrm{T}}}

\title{Butterflies in $\TTbar$ deformed anomalous CFT$_2$}

\author[a,1]{Debarshi Basu\note{Corresponding author.}}
\author[a,2]{and Mingshuai Xu\note{First author.}}

\affiliation[a]{Shing-Tung Yau Center and School of Physics, Southeast University, Nanjing 210096, China}

\emailAdd{debarshi.128@gmail.com, xumingshuai@seu.edu.cn}

\abstract{
	We study quantum chaos in $\TTbar$-deformed two-dimensional conformal field theories with gravitational anomaly and their holographic dual description in topologically massive gravity. Using pole-skipping and shock-wave analysis, we extract the Lyapunov exponent and butterfly velocity and analyze the interplay between irrelevant deformation and parity-violating dynamics. We find that the chaos bound remains saturated, while the butterfly velocity exhibits nontrivial dependence on the deformation parameter and anomaly. We also identify a Hagedorn regime in which the chaotic response becomes complex valued, signaling a breakdown of the physical branch of the deformed theory.
}

\begin{document} 
	\maketitle
	\flushbottom
	
	\section{Introduction}	
	Irrelevant deformations of quantum field theories are typically associated with a loss of predictivity, as they introduce an infinite number of counterterms and uncontrollable ultraviolet behavior. Remarkably, the $\TTbar$ deformation of two-dimensional quantum field theories provides a striking counterexample to this expectation. Originally introduced in the context of integrable models and later understood more generally as a universal deformation generated by the determinant of the stress tensor, the $\TTbar$ operator leads to a class of theories that remain exactly solvable despite being irrelevant \cite{Zamolodchikov:2004ce,Smirnov:2016lqw,Cavaglia:2016oda,Cardy:2018sdv}. In particular, the deformation induces a closed flow equation for the finite-volume spectrum and preserves a surprising degree of analytic control over observables.
	
	A major impetus for the intense recent interest in $\TTbar$ deformations comes from holography. It has been proposed that the deformation admits a precise gravitational interpretation in terms of three-dimensional AdS gravity with modified boundary conditions \cite{McGough:2016lol,Kraus:2018xrn,Guica:2019nzm}. In the simplest picture, the deformation corresponds to placing the dual field theory at a finite radial cutoff in AdS$_3$\footnote{See \cite{Asrat:2017tzd,Shyam:2017znq,Cottrell:2018skz,Hartman:2018tkw,Shyam:2018sro,Jafari:2019qns,Caputa:2019pam,Lewkowycz:2019xse,Giveon:2017myj,Chang:2024voo,Pant:2024eno} for further investigations in this direction.} \cite{McGough:2016lol,Asrat:2017tzd,Taylor:2018xcy,Kraus:2018xrn,Kraus:2022mnu}. Further analyses have clarified that a refined formulation involves mixed boundary conditions for the bulk metric and additional matter fields, thereby providing a consistent variational principle and a precise holographic dictionary \cite{Guica:2019nzm}. In addition to these, several alternative holographic realizations of the $\TTbar$ deformed theory have been proposed in the literature \cite{Dubovsky:2017cnj,Dubovsky:2018bmo,Hirano:2020nwq,Hirano:2020ppu,Apolo:2023ckr,Apolo:2023vnm}.
	
	An important generalization arises when the seed conformal field theory possesses a gravitational anomaly, so that the left- and right-moving central charges are unequal. In this case, the natural bulk dual is provided by topologically massive gravity (TMG), namely three-dimensional Einstein gravity supplemented by a gravitational Chern--Simons term \cite{Deser:1982vy}. TMG is distinguished by its parity-violating structure and admits a rich space of solutions including warped AdS$_3$ geometries and their associated black holes \cite{Anninos:2008fx}.
	
	Recent work has demonstrated that the $\TTbar$ deformation can be consistently extended to anomalous CFT$_2$s and admits a holographic description in terms of TMG with mixed boundary conditions \cite{Basu:2025fsf}. In this framework, one obtains a deformed BTZ black hole geometry together with a modified energy spectrum that satisfies a generalized flow equation incorporating the anomaly. Notably, the deformation preserves many of its universal features, such as solvability and a Hagedorn-like growth of states, while simultaneously introducing new structures associated with chirality and anomaly-induced asymmetry.
	
	A complementary probe of these systems is provided by quantum chaos. In thermal quantum systems with a large number of degrees of freedom, chaos is characterized by the exponential growth of out-of-time-order correlators, governed by the (quantum) Lyapunov exponent and bounded universally by $\lambda_L \leq 2\pi/\beta$ \cite{Maldacena:2015waa}. In holographic theories, this behavior is captured by shock-wave geometries near black hole horizons, which encode the gravitational backreaction of highly boosted perturbations. An equivalent and particularly sharp diagnostic is pole skipping, whereby the retarded Green’s function of the energy-momentum tensor exhibits special points in complex frequency-momentum space at which it is not uniquely defined \cite{Blake:2017ris}. These points are directly related to the Lyapunov exponent and the butterfly velocity, and their existence has been shown to follow from near-horizon dynamics in a wide class of gravitational theories.
	
	The interplay between $\TTbar$ deformation, gravitational anomaly, and quantum chaos is therefore both natural and nontrivial. On one hand, the deformation modifies the effective geometry and thermodynamic quantities of the system, while on the other hand the anomaly introduces a chiral asymmetry that can lead to asymmetric chaotic behavior\footnote{Note that quantum chaos in CFTs dual to TMG has been holographically analysed earlier in \cite{Alishahiha:2016cjk,Liu:2020yaf}.}. From the holographic viewpoint, both effects are encoded in the structure of the bulk equations of motion in TMG and in the choice of boundary conditions. Understanding how these ingredients combine to determine chaotic observables provides valuable insight into the robustness of holographic scrambling and its sensitivity to ultraviolet deformations. A useful bridge to the present discussion is provided by the analysis of rotating shock waves in BTZ geometries\footnote{Interested readers may consult the follow-up works \cite{Malvimat:2022fhd,Malvimat:2022oue,Poojary:2022meo,Poojary:2022vsz,Prihadi:2023qmk,Prihadi:2023tvr,Prihadi:2025czn,Prihadi:2025rwg}.} \cite{Malvimat:2021itk}, where the late-time backreaction of an infalling spinning perturbation leads to a characteristic scrambling of mutual information between the two asymptotic CFTs. In that setting, the relevant growth rate is controlled by an effective Lyapunov exponent $\kappa = \frac{2\pi}{\beta(1-\Omega \mathcal{L})}$, with $\Omega=r_-/r_+$ encoding the background rotation and \(\mathcal{L}\) denoting the angular momentum of the shock wave. For the special case $\mathcal{L}=1$, the mutual information has been evaluated analytically and the scrambling time takes the familiar large-entropy form $t_* \sim \frac{\beta(1-\mu)}{\pi}\log S$. This makes the analysis of rotating shock waves a particularly apt additional step along with the general near-horizon origin of holographic chaos in the deformed, anomalous rotating backgrounds studied in the present work.
	
	In this work, we investigate these questions by studying the chaotic properties of a $\TTbar$-deformed anomalous CFT$_2$ and its dual description in TMG. Our analysis combines several complementary approaches. 
	%First, we review the deformation and its holographic realization in the presence of a gravitational anomaly, emphasizing the role of mixed boundary conditions and the structure of the deformed BTZ background. We then analyze the retarded Green’s functions of the stress tensor and determine the pole-skipping points in the deformed theory, extracting the corresponding Lyapunov exponent and butterfly velocity. Finally, we construct the shock-wave geometry in the rotating black hole background and demonstrate how the deformation and anomaly modify the scrambling dynamics.
	Using the pole-skipping structure of stress-tensor correlators, we determine the chaotic modes associated with the left-moving, right-moving, and massive sectors and extract the corresponding Lyapunov exponents and butterfly velocities. While the Lyapunov exponent saturates the chaos bound in the co-moving frame, the Schwarzschild-frame exponents seem to violate the bound due to the modified causal structure induced by the $\TTbar$ deformation. We argue that this behavior reflects the intrinsically non-local nature of the deformed theory. The butterfly velocities exhibit a nontrivial dependence on both the deformation parameter and the gravitational anomaly, leading to chirally asymmetric information spreading even in the non-rotating ensemble. We further identify a Hagedorn regime in which the chaotic parameters become complex, signaling the breakdown of the physical branch of the deformed theory. We then construct the corresponding shock-wave geometries and analyze the resulting OTOCs, showing that the same chaotic data govern the near-horizon scrambling dynamics. Taking into account the compactness of the spatial direction, we demonstrate that the averaged Lyapunov exponent continues to saturate the maximal chaos bound, paralleling the behavior of the undeformed theory \cite{Mezei:2019dfv,Liu:2020yaf}. We provide a complementary CFT interpretation by mapping the deformed theory to an undeformed CFT on an effective background geometry with renormalized causal structure, thereby clarifying the origin of the modified butterfly cones and superluminal chaotic propagation. Finally, we construct the rotating shock-wave solution in the deformed geometry and show that the fast-scrambling behavior persists, albeit with modified kinematics reflecting the underlying asymmetry and effects of $\TTbar$ deformation.
%	Our main results can be summarized as follows:
%	\begin{itemize}
%		\item We determine the pole-skipping structure of the $\TTbar$-deformed anomalous CFT and show that it encodes the expected chaotic data in a frame adapted to the chiral flow.
%		\item We demonstrate that the Lyapunov exponent continues to saturate the universal chaos bound, while the butterfly velocity exhibits nontrivial dependence on the deformation parameter and the anomaly.
%		\item We construct the rotating shock-wave solution in the deformed geometry and show that the fast-scrambling behavior persists, albeit with modified kinematics reflecting the underlying asymmetry.
%	\end{itemize}
	These results provide a unified picture of how solvable irrelevant deformations and gravitational anomalies affect quantum chaos in holographic systems. More broadly, they illustrate that the near-horizon origin of chaotic behavior remains robust under significant deformations of both the ultraviolet and infrared structure of the theory, while still allowing for rich and physically meaningful modifications.
	
	The remainder of this paper is organized as follows. In section \ref{review}, we review the $\TTbar$ deformation in the presence of a gravitational anomaly and its holographic realization in TMG. Section \ref{sec:Chaos} is devoted to the study of pole skipping, construction of the shock-wave geometry, and the extraction of chaotic observables. In section \ref{Sec.Rotating shockwave}, we analyze fast scrambling in the anomalous deformed background. We conclude in section \ref{sec:summary} with a discussion of implications and future directions.

	\section{Review}\label{review}
	
	\subsection{$\TTbar$ deformation}
	
	The $\TTbar$ deformation is a universal irrelevant deformation of two-dimensional quantum field theory \cite{Zamolodchikov:2004ce,Smirnov:2016lqw,Cavaglia:2016oda,Cardy:2018sdv}, under which the  action in Euclidean signature satisfies the following flow equation\footnote{We follow the conventions in \cite{Guica:2019nzm}.},
	\begin{equation}\label{flow equation}
		\partial_\mu \mathcal{S}^{\left[\mu\right]}=-\frac{1}{2}\int d^2 x  \sqrt{\gamma^{\left[\mu\right]}}\mathcal{O}_{\TTbar}^{\left[\mu\right]},
	\end{equation}
	where $\mu$ is the deformation parameter, and the superscripts denote that the quantities correspond to the deformed theory. The quantity $\gamma_{\alpha\beta}$ is the background metric, and $\mathcal{O}_{\TTbar}$ represents the determinant of the stress tensor,
	\begin{equation}
		\mathcal{O}_{\TTbar}=\gamma_{\alpha\beta}T^{\alpha\eta}T^{\beta}{}_{\eta}-\left(T^{\alpha}{}_{\alpha}\right)^2\,.
	\end{equation} 
	The stress tensor $T_{\alpha\beta}$ is obtained by varying the action with respect to the background metric, i.e.,
	\begin{equation}
		T_{\alpha\beta}=\frac{2}{\sqrt{\gamma}}\frac{\delta \mathcal{S}}{\delta \gamma^{\alpha\beta}}.
	\end{equation}
	As described in \cite{Guica:2019nzm}, through the variational principle, the flow equation \eqref{flow equation} can be reduced to the following differential equations,
	\begin{equation}
		\partial_\mu\gamma^{\left[\mu\right]}_{\alpha\beta}=-2\hat{T}^{\left[\mu\right]}_{\alpha\beta},\quad \partial_\mu\hat{T}_{\alpha\beta}^{\left[\mu\right]}=-\hat{T}^{\left[\mu\right]}_{\alpha\eta}\,\hat{T}^{\left[\mu\right]\eta}{}_{\beta}\,,
	\end{equation}
	where $\hat{T}^{\left[\mu\right]}_{\alpha\beta}$ is the (deformed) trace reversed stress tensor defined as,
	\begin{equation}\label{reversedstd}
		\hat{T}^{\left[\mu\right]}_{\alpha\beta}\equiv T_{\alpha\beta}^{\left[\mu\right]}-\gamma_{\alpha\beta}^{\left[\mu\right]}T^{\left[\mu\right]\eta}{}_{\eta}.
	\end{equation}
	The non-perturbative analytical solutions can be obtained in terms of the undeformed metric and stress tensor $(\gamma_{\alpha\beta}^{[0]}, \hat{T}_{\alpha\beta}^{[0]})$,
	\begin{align}
		\gamma^{\left[\mu\right]}_{\alpha\beta}=&\:\gamma_{\alpha\beta}^{\left[0\right]}-2\mu\,\hat{T}_{\alpha\beta}^{\left[0\right]}+\mu^2\hat{T}_{\alpha\rho}^{\left[0\right]}\,\hat{T}_{\sigma\beta}^{\left[0\right]}\,\gamma^{\left[0\right]\rho\sigma},\label{flowsolutions}\\
		\hat{T}^{\left[\mu\right]}_{\alpha\beta}=&\:\hat{T}_{\alpha\beta}^{\left[0\right]}-\mu\,\hat{T}_{\alpha\rho}^{\left[0\right]}\,\hat{T}^{\left[0\right]}_{\sigma\beta}\,\gamma^{\left[0\right]\rho\sigma}.\label{deformedst}
	\end{align}
	In particular, if we consider the seed theory to be a conformal field theory, the deformed energy spectrum may be obtained from both field theory and bulk sides, and is given by,
	\begin{equation}
		E_\mu=\frac{R}{2\mu}\left(-1+\sqrt{1+4\mu\left(\frac{E}{R}+\mu\frac{J^2}{R^2}\right)}\right),
	\end{equation}
	where $E$ and $J$ are the undeformed energy and the angular momentum respectively, and $R$ is the period of the spatial coordinate. And we used a rescaled deformation parameter $\mu \rightarrow \mu/8\pi G$ for simplicity and we keep this convention throughout this paper, unless specified otherwise.

	\subsubsection{TMG with $\TTbar$ deformation}

	In this paper, the seed field theory we focus on is the two dimensional conformal field theory with gravitational anomalies \cite{Alvarez-Gaume:1983ihn,Alvarez-Gaume:1984zlq}, which is characterized by two unequal left-moving and right-moving central charges and is denoted as CFT$_2^a$. The gravity dual is proposed to be topological massive gravity (TMG), whose action includes the usual the Einstein-Hilbert term together with a gravitational Chern-Simons (CS) term \cite{Deser:1982vy}
	\begin{equation}
		I_{\text{TMG}}=\frac{1}{16\pi G_N}\int d^3x\sqrt{-g}\left(R-2\Lambda +\frac{1}{2\lambda}\varepsilon^{\alpha\beta\gamma}\left(\Gamma^{\rho}{}_{\alpha\sigma}\partial_\beta \Gamma^{\sigma}{}_{\gamma\rho}+\frac{2}{3}\Gamma^{\rho}{}_{\alpha\sigma}\Gamma^{\sigma}{}_{\beta\eta}\Gamma^{\eta}{}_{\gamma\rho}\right)\right),
	\end{equation}
	where $\Lambda=-1$ is the cosmological constant\footnote{We set the AdS radius $\ell = 1$ in this paper.} and $\lambda$ is the coupling constant that characterizes the interaction strength between the CS term and the Einstein-Hilbert term. The equations of motion of TMG are given by \cite{Deser:1982vy},
	\begin{align}\label{TMG-EoM}
		R_{\mu\nu}-\frac{1}{2}g_{\mu\nu}R+\Lambda g_{\mu\nu}=-\frac{1}{\lambda} C_{\mu\nu}
	\end{align}
	where $C_{\mu\nu}$ is the Cotton tensor, defined as
	\begin{align}
		C_{\mu\nu}=\varepsilon_\mu^{~\,\alpha\beta}\,\nabla_\alpha \CS_{\beta\nu}~~,~~\CS_{\mu\nu}=R_{\mu\nu}-\frac{1}{4}g_{\mu\nu}R\,.
	\end{align}
	The equations of motion are third order in derivatives, hence the solutions of the field equations require the specification of the metric and its derivatives at the boundary. On the other hand, for locally AdS$_3$ geometries the Cotton tensor vanishes identically, and therefore, the solutions of Einstein gravity are solutions to TMG as well. The Brown-Henneaux symmetry analysis \cite{Brown:1986nw} of TMG on locally AdS$_3$ backgrounds shows that there are two unequal central charges in the asymptotic symmetry algebra \cite{Kraus:2005zm,Hotta:2008yq,Compere:2008cv},
	\begin{equation}\label{twocentreal}
		c_L=\frac{3}{2G}\left(1+\frac{1}{\lambda}\right), \quad c_R=\frac{3}{2G}\left(1-\frac{1}{\lambda}\right).
	\end{equation}
	Since the central charges characterize the degrees of freedom of the system, it is natural to require that both $c^{}_L$ and $c^{}_R$ are positive, i.e., $\lambda < -1$ or $\lambda > 1$.
	
	Now we turn to the holographic description of the $\TTbar$ deformed gravitational anomalous CFT$_2$ developed in \cite{Basu:2025fsf}. There are mainly two holographic descriptions for the $\TTbar$ deformation. The first one, dubbed the cutoff prescription, introduces the Dirichlet boundary conditions on a specific radial slice, on which the induced metric should align with the deformed background metric of the boundary field theory, up to a global constant conformal factor \cite{McGough:2016lol,Asrat:2017tzd,Taylor:2018xcy,Kraus:2018xrn,Kraus:2022mnu}. The second one is a more ﬂexible approach: the mixed (Dirichlet-Neumann) boundary condition prescription, which involves the deformed bulk geometry through introducing mixed nonlinear boundary conditions \cite{Guica:2019nzm}. However, it was shown in \cite{Tian:2024vln} that the deformed boundary background metric can not be understood as the induced metric on a specific bulk radial slice, hence the cutoff prescription does not work for the TMG. Consequently, \cite{Basu:2025fsf} adopted the mixed boundary condition prescription to study the holographic description of the $\TTbar$ deformed gravitational anomalous CFT$_2$, which is briefly reviewed in the following. We start with an auxiliary rotating BTZ black hole geometry in the radial Fefferman-Graham gauge\footnote{In fact, we can consider a more general situation where $\mathcal{L}_\mu$ and $\bar{\mathcal{L}}_\mu$ are two arbitrary chiral functions, corresponding to the Ba\~{n}ados geometry that describes general AdS$_3$ solutions with a flat boundary. However, in this paper we focus on the rotating BTZ black hole, which is dual to a CFT at finite temperature with a conserved angular momentum.},
	\begin{equation}\label{auxiliary geometry}
		ds^2 =\frac{d\rho^2}{4\rho^2} + \frac{dw d\bar{w}}{\rho} + \mathcal{L}_\mu dw^2 + \bar{\mathcal{L}}_\mu d\bar{w}^2 + \rho \mathcal{L}_\mu \bar{\mathcal{L}}_\mu dwd\bar{w},
	\end{equation}
	where the parameters $\left(\mathcal{L}_\mu,\bar{\mathcal{L}}_\mu\right)$ are two constants, and $\left(w,\bar{w}\right)$ are two boundary null coordinates, defined as $w,\bar{w}=\varphi\pm \tau$. This line element can be written in the standard ADM form,
	\begin{equation}\label{BTZgeometry}
		ds^2=-\frac{\left(r^2-r_+^2\right)\left(r^2-r_-^2\right)}{r^2}d\tau^2+\frac{r^2}{\left(r^2-r_+^2\right)\left(r^2-r_-^2\right)}dr^2+r^2\left(d\varphi-\frac{r_+r_-}{r^2}d\tau\right)^2,
	\end{equation} 
	by the following radial coordinate transformation,
	\begin{equation}\label{radialct}
		r=\sqrt{\frac{r_+^2+r_-^2}{2}+\frac{1}{\rho}+\frac{\rho}{16}r_h^4},\quad r_h=\sqrt{r_+^2-r_-^2}.
	\end{equation}
	The relations between the parameters $\left(\mathcal{L}_\mu,\bar{\mathcal{L}}_\mu\right)$ and the radii of the horizons $\left(r_+,r_-\right)$ read,
	\begin{equation}\label{r2r1Lmu}
		r_+=\sqrt{\bar{\mathcal{L}}_\mu}+\sqrt{\mathcal{L}_\mu},\quad r_-=\sqrt{\bar{\mathcal{L}}_\mu}-\sqrt{\mathcal{L}_\mu}.
	\end{equation}
	According to the holographic renormalization \cite{Skenderis:1999nb}, the holographic stress tensor of TMG will receive corrections from the CS term correspondingly \cite{Skenderis:2009nt}, i.e.,
	\begin{equation}\label{ustTMG}
		\hat{T}_{\alpha\beta}^{\left[0\right]}\left(w,\bar{w}\right)=T_{\alpha\beta}^{\left[0\right]}\left(w,\bar{w}\right)=\frac{1}{8\pi G}\text{diag}\left(\kappa\mathcal{L}_\mu,\bar{\kappa}\bar{\mathcal{L}}_\mu\right),
	\end{equation}
	where $\kappa,\bar{\kappa}=1\pm1/\lambda$, and we have used the undeformed background $\gamma^{\left[0\right]}_{\alpha\beta}dx^\alpha dx^\beta=dwd\bar{w}$ and the definition of the trace-reversed stress tensor \eqref{reversedstd}. Substituting this expression into \eqref{flowsolutions}, we can obtain the field-dependent coordinate transformations by imposing the mixed boundary conditions,
	\begin{equation}\label{deformedboume}
		\gamma^{\left[\mu\right]}_{\alpha\beta}\:dx^\alpha dx^\beta=\left(dw-2\mu\bar{\kappa}\bar{\mathcal{L}}_\mu d\bar{w}\right)\left(d\bar{w}-2\mu\kappa \mathcal{L}_\mu dw\right)=dZd\bar{Z},
	\end{equation}
	which implies,
	\begin{equation}\label{coordinatetrans}
		Z=w-2\mu\bar{\kappa}\bar{\mathcal{L}}_\mu \bar{w},\qquad \bar{Z}=\bar{w}-2\mu\kappa\mathcal{L}_\mu w.
	\end{equation}
	%	\begin{equation}\label{coordinatetrans1}
		%		 u=\frac{U+2\mu\bar{\kappa}\bar{\mathcal{L}}_\mu V}{1-4\mu^2\kappa\bar{\kappa}\mathcal{L}_\mu\bar{\mathcal{L}}_\mu},\qquad v=\frac{V+2\mu\kappa\mathcal{L}_\mu U}{1-4\mu^2\kappa\bar{\kappa}\mathcal{L}_\mu\bar{\mathcal{L}}_\mu}.
		%	\end{equation}
	Plugging the field-dependent coordinate transformations back into \eqref{auxiliary geometry}, we can obtain the deformed bulk metric \cite{Basu:2025fsf},
	\begin{align}
		\d s^2=\frac{\d\rho^2}{4\rho^2}&+\left(\frac{1}{\rho}+\rho\CL_\mu\bar\CL_\mu\right)\frac{\left(\d Z+2\mu\bar\kappa\bar\CL_\mu \d\bar Z\right)\left(\d \bar Z+2\mu\kappa\CL_\mu \d Z\right)}{\left(1-4\mu^2\kappa\bar\kappa\CL_\mu\bar\CL_\mu\right)^2}\notag\\
		&+\frac{\CL_\mu\left(\d Z+2\mu\bar\kappa\bar\CL_\mu \d\bar Z\right)^2+\bar\CL_\mu\left(\d \bar Z+2\mu\kappa\CL_\mu \d Z\right)^2}{\left(1-4\mu^2\kappa\bar\kappa\CL_\mu\bar\CL_\mu\right)^2}.
	\end{align}
	With the deformed bulk metric, we can also discuss the deformed energy spectrum and the Hagedorn behavior of the $\TTbar$ deformed anomalous CFT, see \cite{Basu:2025fsf} for details. Using the radial coordinate transformation \eqref{radialct}, one can write down the above $\TTbar$ deformed bulk metric in the Schwarzschild coordinate system,
	\begin{align}\label{deformed metric}
		\d s^2=\frac{r^2\d r^2}{(r^2-r_+^2)(r^2-r_-^2)}+\frac{1}{r_h^2\alpha_\mu^2}&\Big[-(r^2-r_+^2)\left(r_+(1-\sigma_\mu)\d t-r_-(1-\nu_\mu)\d x\right)^2\notag\\
		&+(r^2-r_-^2)\left(r_-(1+\nu_\mu)\d t-r_+(1+\sigma_\mu)\d x\right)^2\Big]\,,
	\end{align}
	where $Z,\bar{Z}=x\pm t$, and we have used the following shorthand notations,
	\begin{align}\label{sigma-nu}
		\sigma_\mu=\frac{r_h^2}{2}\left(1-\frac{r_-}{r_+\lambda}\right)\mu,~~\nu_\mu=\frac{r_h^2}{2}\left(1-\frac{r_+}{r_-\lambda}\right)\mu,~~\alpha_\mu=\frac{r_+^2\left(1-\sigma_\mu^2\right)-r_-^2\left(1-\nu_\mu^2\right)}{r_h^2}.
	\end{align}
	Note that in the absence of gravitational anomaly, we have $\sigma_\mu=\nu_\mu=\frac{1}{2}r_h^2\mu$.
	The thermal identification of the $\TTbar$ deformed system reads,
	\begin{align}
		(Z,\bar Z)\sim (Z+i\beta_+,\bar Z-i\beta_-),
	\end{align}
	with the left and right-moving inverse temperatures $\left(\beta_+,\beta_-\right)$ expressed as \cite{Basu:2025fsf}
	\begin{equation}\label{beta-pm}
		\begin{aligned}
			\beta_+=&\frac{\pi}{\sqrt{\CL_\mu}}+2\pi\mu\bar\kappa\sqrt{\bar\CL_\mu}=\frac{2\pi}{r_+-r_-}+\pi\mu\bar{\kappa}\left(r_++r_-\right),\\\beta_-=&\frac{\pi}{\sqrt{\bar\CL_\mu}}+2\pi\mu\kappa\sqrt{\CL_\mu}=\frac{2\pi}{r_++r_-}+\pi\mu\kappa\left(r_+-r_-\right),
		\end{aligned}
	\end{equation}
	where we have used \eqref{r2r1Lmu}. In the Schwarzschild coordinate system, one has,
	\begin{equation}
		\left(x,t\right)\sim\left(x+i\beta\Omega,t+i\beta\right),
	\end{equation}
	where $\beta$ and $\Omega$ denote the inverse temperature and the angular speed respectively, with the explicit forms,
	\begin{equation}\label{beta-Omega}
		\beta=\frac{\beta_++\beta_-}{2}=\frac{2\pi r_+\left(1+\sigma_\mu\right)}{r_h^2},\quad \Omega=\frac{\beta_+-\beta_-}{\beta_++\beta_-}=\frac{r_-\left(1+\nu_\mu\right)}{r_+\left(1+\sigma_\mu\right)}.
	\end{equation}
	The relations between the radii of horizons $\left(r_+,r_-\right)$ and $\left(\beta,\Omega\right)$ are given by,
	\begin{equation}\label{r-pm}
		\begin{aligned}
			r_+&=\frac{2 \pi  (1+\lambda  \,\Omega)}{\beta  \left(1-\lambda ^2\right) \left(1-\Omega ^2\right)}-\frac{\beta  \lambda  (\lambda +\Omega )}{2 \pi  \left(1-\lambda ^2\right) \mu}\left(1-\sqrt{1-\frac{8 \pi ^2 \mu }{\beta _+ \beta _-}+\left(\frac{4\pi^2\mu}{\beta_+\beta_-\lambda}\right)^2}\right)\,,\\
			r_-&=\frac{2 \pi  (\lambda +\Omega )}{\beta  \left(1-\lambda ^2\right) \left(1-\Omega ^2\right)}-\frac{\beta  \lambda  (1+\lambda \,\Omega)}{2 \pi  \left(1-\lambda ^2\right) \mu}\left(1-\sqrt{1-\frac{8 \pi ^2 \mu }{\beta _+ \beta _-}+\left(\frac{4\pi^2\mu}{\beta_+\beta_-\lambda}\right)^2}\right)\,.
		\end{aligned}
	\end{equation}

	\subsection{Pole skipping}
	A particularly powerful probe of quantum chaos in holographic systems is provided by the analytic structure of retarded Green's functions. In thermal states, the response of the system to small perturbations is encoded in the poles of these Green's functions, which determine the spectrum of collective excitations. Remarkably, in holographic theories dual to black hole geometries, the retarded Green's function of energy-density operators exhibits a characteristic phenomenon known as \emph{pole skipping} \cite{Grozdanov:2017ajz,Blake:2017ris,Blake:2018leo,Grozdanov:2018kkt}. This phenomenon reflects the breakdown of the usual pole structure at special complex values of frequency and momentum, and provides a direct window into chaotic dynamics.
	
	Concretely, one considers the retarded correlator of the energy-momentum tensor, which may be written schematically as
	\begin{equation}
		G^{R}_{T^{00}T^{00}}(\omega,k)=\frac{A(\omega,k)}{B(\omega,k)}\,.
	\end{equation}
	At generic values of $(\omega,k)$, the poles of the Green’s function are determined by the zeroes of $B(\omega,k)$. However, at special points $(\omega_\star,k_\star)$, both the numerator and denominator vanish simultaneously \cite{Blake:2019otz},
	\begin{equation}
		A(\omega_\star,k_\star)=0\,, \qquad B(\omega_\star,k_\star)=0\,,
	\end{equation}
	and the Green’s function becomes ill-defined. These points are the pole-skipping locations, and they encode the fundamental chaotic data of the system,
	\begin{equation}
		\omega_\star = i \lambda_L\,, \qquad k_\star = i \frac{\lambda_L}{v_B}\,,\label{Pole-skipping-point}
	\end{equation}
	where $\lambda_L$ is the Lyapunov exponent and $v_B$ is the butterfly velocity.
	
	From the holographic perspective, pole skipping arises from a degeneracy in the near-horizon behavior of bulk perturbations. Working in the infalling Eddington--Finkelstein coordinates $(\mathbf{v},r,\phi)$, one analyzes linearized gravitational fluctuations with harmonic dependence $e^{-i(\omega \mathbf{v} - k \phi)}$. Near the horizon, the equations of motion reduce to a set of recursion relations for the expansion coefficients of the perturbations. At generic values of $(\omega,k)$, these relations uniquely determine the solution in terms of boundary data. However, at the pole-skipping points, the leading constraint degenerates, and the system admits an additional free parameter. This degeneracy is the bulk manifestation of the non-uniqueness of the Green’s function and is directly tied to the exponential sensitivity characteristic of chaotic systems.
	
	In the present context, the analysis is further enriched by the presence of both a $\TTbar$ deformation and a gravitational anomaly. The deformation modifies the background geometry and thermodynamic quantities, while the anomaly introduces parity-violating contributions through the Cotton tensor in TMG. Nevertheless, the pole-skipping mechanism remains governed by near-horizon physics, and the resulting Lyapunov exponent continues to be determined by the effective surface gravity of the deformed black hole. The butterfly velocity, on the other hand, acquires nontrivial dependence on both the deformation parameter and the chiral structure induced by the anomaly.
	%--------------------------------------------------------------------------------------------------------------------------------
	\subsection{Shockwave method and OTOC}
	
	The pole-skipping phenomenon admits a complementary and physically intuitive interpretation in terms of shock-wave geometries in the bulk. In this approach, one considers a localized perturbation created by an operator insertion at an early boundary time $t_w$, which propagates into the bulk and becomes highly boosted as it approaches the black hole horizon \cite{Sekino:2008he,Shenker:2013pqa,Shenker:2014cwa,Roberts:2014ifa,Roberts:2014isa,Roberts:2016wdl}. The backreaction of this perturbation produces a gravitational shock wave localized on a null surface, modifying the near-horizon geometry in a way that captures the onset of chaotic behavior.
	
	This picture is directly connected to the behavior of out-of-time-order correlators (OTOCs), which provide a standard diagnostic of quantum chaos \cite{Maldacena:2015waa,Shenker:2013pqa}. For two generic operators $V$ and $W$, the thermal OTOC takes the form
	\begin{equation}
		C(t,\phi) = \langle V(t,\phi)\, W(0,0)\, V(t,\phi)\, W(0,0) \rangle_\beta\,,\label{OTOC-defn}
	\end{equation}
	and its deviation from factorization measures the growth of operator commutators. In chaotic systems, this deviation grows exponentially at early times\footnote{See, e.g., \cite{Xu:2018dfp, Gharibyan:2018fax, Steinberg:2019uqb, Gu:2021xaj, Poojary:2018esz, Banerjee:2018twd, Malvimat:2021itk, Das:2022jrr, Das:2021qsd, Das:2019tga,Fan:2016ean,Biswas:2024mlq,Baishya:2024sym,Banerjee:2022ime,Balasubramanian:2019stt,Craps:2020ahu,Chakrabortty:2022kvq,Xu:2022vko,Das:2022pez,Hashimoto:2020xfr,Khetrapal:2022dzy,Perlmutter:2016pkf,Saha:2024bpt} for further studies of OTOCs in various systems.},
	\begin{equation}
		C(t,\phi) \sim 1 - \epsilon^{}_{VW}\, e^{\lambda_L \left(t -t_\star- \frac{|\phi|}{v_B}\right)}\,,
	\end{equation}
	where $\lambda_L$ is the Lyapunov exponent, $t_\star$ is termed the scrambling time and $v_B$ is the butterfly velocity. In holography, this exponential behavior is reproduced by the eikonal scattering of bulk particles near the horizon, which is precisely captured by the shock-wave geometry. In Kruskal coordinates, the effect of the shock wave is to introduce a shift in one of the null coordinates across the horizon,
	\begin{equation}
		V \rightarrow V + \alpha\, f(\phi)\,,
	\end{equation}
	where $f(\phi)$ is the transverse shock profile and $\alpha \sim e^{\lambda_L t_w}$ encodes the exponential growth of the perturbation with the insertion time $-t_w$. The profile function satisfies a differential equation determined by the bulk equations of motion,
	\begin{equation}
		\CD f(\phi) \sim \alpha\delta(\phi)\,,
	\end{equation}
	where $\CD$ is a differential operator acting along the horizon. The resulting shift modifies the geodesic distance between boundary points and hence the two-point functions entering the OTOC. As a result, the OTOC acquires an exponential suppression governed by the same Lyapunov exponent.
	
	In Einstein gravity, the operator $\CD$ is typically second order, but in topologically massive gravity it becomes third order due to the presence of the gravitational Chern--Simons term \cite{Alishahiha:2016cjk,Liu:2020yaf}. This modification reflects the parity-violating nature of the theory and leads to a chiral structure in the propagation of perturbations. Consequently, the OTOC can exhibit directional dependence, with different effective butterfly velocities for left- and right-moving modes \cite{Liu:2020yaf}. This provides a direct bulk realization of the asymmetry induced by the gravitational anomaly in the boundary theory.
	
	While pole skipping identifies the chaotic point through the analytic structure of the retarded Green’s function, the shock-wave method derives the same exponential behavior from the bulk scattering amplitude that governs the OTOC. Their agreement highlights the universal role of near-horizon physics in governing chaos. Taken together, these results establish a unified picture in which the exponential growth of OTOCs, the pole-skipping phenomenon, and the shock-wave backreaction are different manifestations of the same underlying mechanism \cite{Chua:2025vig,Dong:2022ucb}. In the presence of a $\TTbar$ deformation and a gravitational anomaly, this mechanism remains intact, although the detailed values of the butterfly velocity and the spatial profile of the perturbation are modified. In the following sections, we will apply this framework to the deformed anomalous system and compute the corresponding chaotic observables explicitly.
	
	\section{Quantum chaos in TMG with $\TTbar$ deformation}\label{sec:Chaos}
	
	In this section, we investigate the chaotic properties of the $\TTbar$ deformed BTZ black hole \eqref{deformed metric} in the presence of gravitational anomaly.
	In the Schwarzschild coordinate system, the inner and outer horizons are at $r=r_\pm$ which are related to the left and right moving temperatures $\beta_\pm$ through the relations \eqref{beta-pm} and \eqref{r-pm}. The coordinates $(t,x)$ align with the boundary time and spatial coordinates, hence we focus on the Lyapunov growth of chaos and information spreading in these canonical coordinates in the following.
	%------------------------------------------------------------------------------
	\paragraph{Co-moving coordinates:} Our aim is to find a spatial coordinate with no periodic identification, in which the effects of the black hole spin are eliminated. A canonical choice for the co-moving coordinate is given as
	\begin{align}
		\phi=x-\frac{\beta_+-\beta_-}{\beta_++\beta_-}t\equiv x-\Omega t\label{co-moving}
	\end{align}
	where the angular velocity (at the horizon) $\Omega$ of the deformed spacetime is given in \eqref{beta-Omega}.
	Note that, in the absence of deformation, $\sigma_\mu=\nu_\mu=0$ and the angular velocity $\Omega$ reduces to the well-known expression $\Omega(\mu=0)=\frac{r_-}{r_+}$. In the co-moving coordinates, the spacetime metric takes the form
	\begin{align}
		\d s^2=&\frac{r^2}{\left(r^2-r_+^2\right)\left(r^2-r_-^2\right)}\d r^2-\frac{\left(r^2-r_+^2\right)r_h^2}{r_+^2\left(1+\sigma_\mu\right)^2}\d t^2+\frac{2r_-\left(r^2-r_+^2\right)\left(1-\nu_\mu\right)}{r_+\alpha_\mu\left(1+\sigma_\mu\right)}\d t\,\d\phi\notag\\
		&+\frac{1}{r_h^2\alpha_\mu^2}\left(r_+^2\left(1+\sigma_\mu\right)^2\left(r^2-r_-^2\right)-r_-^2\left(1-\nu_\mu\right)^2\left(r^2-r_+^2\right)\right)\d\phi^2\,.\label{Co-moving-metric}
	\end{align}
	Furthermore, the frequencies and momenta in the co-moving and Schwarzschild ccordinates are related through the inverse transformations of \eqref{co-moving} as follows \cite{Liu:2020yaf},
	\begin{align}
		\omega_\textrm{Sch}=\omega_\textrm{cm}+\Omega\, k_{cm}~~,~~k_\textrm{Sch}=k_\textrm{cm}\,.\label{frequency-transformations}
	\end{align}
	
	\subsection{Pole skipping}\label{Poleskipping}
	
	In this subsection, we aim to holographically analyze the retarded energy-density Green's function $G^R_{T^{00}T^{00}}(\omega,k)$ in the deformed anomalous CFT$_2$.
	It is customary to work in the infalling Eddington-Finkelstein (EF) coordinates $(\mathbf{v},r,\phi)$, with\footnote{Note that, we do not need to keep track of the transformation to the infalling EF coordinates to recover the frequencies in the Schwarzschild coordinates, since the advanced EF time $\mathbf{v}$ reduces to the Schwarzschild time $t$ when approaching the asymptotic boundary $r_\star\to 0$.}
	\begin{align}
		\mathbf{v}=t+r_\star\,,
	\end{align}
	where the tortoise coordinate $r_\star$ can be derived from the co-moving metric as follows
	\begin{align}\label{rstar}
		\d r_\star=\frac{r\,r_+(1+\sigma_\mu)}{r_h(r^2-r_+^2)\sqrt{r^2-r_-^2}}\d r\implies r_\star=\frac{r_+(1+\sigma_\mu)}{2r_h^2}\log\left(\frac{\sqrt{r^2-r_-^2}-r_h}{\sqrt{r^2-r_-^2}+r_h}\right)
	\end{align}
	In these coordinates, the deformed metric finally takes the form
	\begin{align}
		\d s^2=&-\frac{r_h^2\left(r^2-r_+^2\right)\d\mathbf{v}^2}{r_+^2\left(1+\sigma_\mu\right)^2}+\frac{2r r_h\d\mathbf{v}\,\d r}{r_+\left(1+\sigma_\mu\right)\sqrt{r^2-r_-^2}}+\frac{2r_-\left(r^2-r_+^2\right)\left(1-\nu_\mu\right)\d\mathbf{v}\,\d\phi}{r_+\alpha_\mu\left(1+\sigma_\mu\right)}\notag\\
		&-\frac{2r\,r_-\left(1-\nu_\mu\right)\d r\,\d\phi}{r_h\alpha_\mu\sqrt{r^2-r_-^2}}+\frac{1}{r_h^2\alpha_\mu^2}\left(r_+^2\left(1+\sigma_\mu\right)^2\left(r^2-r_-^2\right)-r_-^2\left(1-\nu_\mu\right)^2\left(r^2-r_+^2\right)\right)\d\phi^2\,.\label{infalling-EF}
	\end{align}
	It is easy to verify that the deformed BTZ black hole metric in infalling EF coordinates given in \eqref{infalling-EF}, satisfies the equations of motion \eqref{TMG-EoM}. We then proceed to analyze the pole skipping phenomena for the near-horizon infalling acoustic modes. To obtain the energy density Green's function in the momentum space, we now perturb the background metric by introducing perturbations of the longitudinal modes or the sound modes. In particular, we adopt the following ansatz \cite{Blake:2019otz,Blake:2018leo},
	\begin{align}
		g_{\mu\nu}\to g_{\mu\nu}+h_{\mu\nu}=g_{\mu\nu}+\delta g_{\mu\nu}(r)\,e^{-i\left(\omega \mathbf{v}-k \phi\right)}\,,\label{logitudinal-perturbations}
	\end{align}
	where $g_{\mu\nu}$ satisfies the vacuum TMG equations of motion. 
	Plugging the above ansatz into \eqref{TMG-EoM}, we may obtain the field equations for $h_{\mu\nu}$ as follows
	\begin{align}
		\delta E_{\mu\nu}=-\frac{1}{\lambda}\delta C_{\mu\nu}\,.\label{Linearized-EoM}
	\end{align}
	The linearized Einstein tensor and Cotton tensor are given by
	\begin{align}
		\delta E_{\mu\nu}=\delta R_{\mu\nu}-\frac{1}{2}g_{\mu\nu}\delta R+2h_{\mu\nu}~~,~~\delta C_{\mu\nu}=\varepsilon_\mu^{~\,\alpha\beta}\nabla_\alpha\left(\delta S_{\beta\nu}+\frac{1}{2}h_{\beta\nu}\right)\label{Linear-EH-Cotton}
	\end{align}
	with
	\begin{align}
		\delta R_{\mu\nu}&=\nabla^{\alpha}\nabla_{(\mu}h_{\nu)\alpha}-\frac{1}{2}\nabla^\alpha\nabla_\alpha h_{\mu\nu}-\frac{1}{2}\nabla_\mu\nabla_\nu h\,,\notag\\
		\delta R&=\nabla^\mu\nabla^\nu h_{\mu\nu}-\nabla^2h+2h\,,\notag\\
		\delta S_{\mu\nu}&=\delta R_{\mu\nu}-\frac{1}{4}g_{\mu\nu}\delta R-\frac{1}{4}Rh_{\mu\nu}\,,\label{linearized-tensors}
	\end{align}
	and $h=g^{\mu\nu}h_{\mu\nu}$. The above equations can also be obtained from the second order expansion of the TMG action around the background \eqref{infalling-EF}.
	The relevant sound-mode metric perturbations consist of
	\begin{align*}
		\delta g_{\mathbf{v}\mathbf{v}}\,,\,\delta g_{ \mathbf{v} \phi}\,,\,\delta g_{\mathbf{v} r}\,,\,\delta g_{rr}\,,\,\delta g_{r\phi}\,,\,\delta g_{\phi\phi}\,.
	\end{align*}
	Imposing the radial gauge condition $\delta g_{r\mu}=0$, the non-redundant modes are only 
	\begin{align*}
		\delta g_{\mathbf{v}\mathbf{v}}\,,\,\delta g_{\mathbf{v}\phi}\,,\,\delta g_{\phi\phi}\,.
	\end{align*}
	Furthermore, the perturbed equations should be regular at the horizon in the infalling EF coordinates. We now allow for the perturbations on the longitudinal modes and expand the fields near the (outer) horizon\footnote{Note that, the Einstein's equations are regular at the inner horizon.}
	\begin{align}
		\delta g_{\mu\nu}(r)=\sum_{n=0}^{\infty}\delta g_{\mu\nu}^{(n)}(r-r_+)^n\,.\label{expanding-near-rp}
	\end{align}
	As described in \cite{Blake:2018leo,Wang:2022mcq}, the near horizon expansion of the linearized Einstein's equation \eqref{Linearized-EoM} becomes degenerate at the special points \eqref{Pole-skipping-point}, admitting an additional infalling mode and causing coincident zeros and poles in the retarded Green's function of the energy‑density correlator \cite{Blake:2019otz}. This directly relates the chaos parameters, the Lyapunov exponent $\lambda_L$ and butterfly velocity $v_B$, to the analytic structure of thermal two‑point functions. This feature has been verified in both anisotropic plasma models \cite{Grozdanov:2018kkt} and massive gravity backgrounds \cite{Ceplak:2021efc}. Accordingly, the Einstein equations, along with their generalizations to other bulk fields, provide a universal gravitational origin for quantum chaos within holography. We may expand the Einstein's equations near the horizon for the perturbed sound modes \eqref{logitudinal-perturbations}. Expanding the linearized Einstein tensor \eqref{Linear-EH-Cotton} near the horizon, we obtain
	\begin{align}
		\delta E_{\mathbf{v}\mathbf{v}}=&\frac{e^{i (k \phi -\omega \mathbf{v})}}{2 r_h^2 r_+^3 (1+\sigma_\mu )^3}\Bigg(\alpha_\mu^2 r_h^2 (2 \delta g_{\textbf{v}\phi}^{(0)} k+\delta g_{\phi\phi}^{(0)} \omega ) \left(r_+ (1+\sigma_\mu) \omega -i r_h^2\right)+\delta g_{\textbf{v}\textbf{v}}^{(0)} r_+ (1+\sigma_\mu) \notag\\&\times \Big(k\alpha_\mu r_h^2\left(k\alpha_\mu-2i r_-\left(1-\nu_\mu\right)\right) -i r_+ (1+\sigma_\mu) \omega  \left(r_+^2 (1+\sigma_\mu)^2-r_-^2 (1-\nu_\mu)^2 \right)\Big)\Bigg)\label{delta-Einstein}
	\end{align}
	For now, we neglect the contribution from the Cotton tensor and analyze the degenerate behavior of the linearized Einstein tensor $\delta E_{\mu\nu}$ near the horizon at special values of $(\omega,k)$. We now analytically continue the freqency and mementum of the infalling metric perturbations to the complex plane. Equating the coefficient of $\delta g_{\mathbf{v}\phi}$ to zero, we may obtain the critical value of the frequency,
	\begin{align}
		\omega_\star=i\frac{r_h^2}{r_+ \left(1+\sigma _{\mu }\right)}=\frac{2\pi i}{\beta}\,.\label{omega-critical}
	\end{align}
	where in the second equality, we have inverted the relations \eqref{sigma-nu} using \eqref{r-pm} and \eqref{beta-pm}. Now substituting the above expression for $\omega_\star$, we may simplify the coefficient of $\delta g_{\mathbf{v}\mathbf{v}}$ in \eqref{delta-Einstein} as follows
	\begin{align}
		\frac{1}{2r_+^2\left(1+\sigma_\mu\right)^2}\left(r_+^2 (1+\sigma_\mu)^2+(\alpha_\mu  k-i r_-(1-\nu_\mu))^2\right)=0\,.
	\end{align}
	This leads to two solutions for $k_\star$:
	\begin{align}\label{kstartwo}
		k_\star&=\frac{i\left(r_-(1-\nu_\mu)\pm r_+(1+\sigma_\mu)\right)}{\alpha_\mu}=\frac{2 \pi i}{\beta(1-\Omega^2)}\left(\Omega \pm \frac{1\pm \frac{4 \pi ^2 \mu }{\beta _+ \beta _- \lambda }}{\sqrt{1-\frac{8 \pi ^2 \mu }{\beta _+ \beta _-}+\left(\frac{4\pi^2\mu}{\beta_+\beta_-\lambda}\right)^2}}\right)
	\end{align}
	
	Now we consider the near horizon expansion of the linearized Cotton tensor in \eqref{Linear-EH-Cotton} and \eqref{linearized-tensors} to obtain
	\begin{align}
		\delta C_{\mathbf{v}\mathbf{v}}=\frac{e^{i (k \phi -\omega \mathbf{v})}}{2 r_h^4 r_+^4 (1+\sigma_\mu)^4}\Big(c_{\mathbf{v}\mathbf{v}}^{(0)}\delta g_{\mathbf{v}\mathbf{v}}^{(0)}+c_{\mathbf{v}\phi}^{(0)}\delta g_{\mathbf{v}\phi}^{(0)}+c_{\phi\phi}^{(0)}\delta g_{\phi\phi}^{(0)}+c_{\mathbf{v}\mathbf{v}}^{(1)}\delta g_{\mathbf{v}\mathbf{v}}^{(1)}+c_{\mathbf{v}\phi}^{(1)}\delta g_{\mathbf{v}\phi}^{(1)}+c_{\phi\phi}^{(1)}\delta g_{\phi\phi}^{(1)}\Big)\label{linearized-Cotton}
	\end{align}
	where the coefficients are listed in appendix \ref{AppA}. We next analyze the breakdown of the full  linearized TMG equations of motion \eqref{Linearized-EoM} at the special values of $(\omega,k)$. Remarkably, we find that the coefficients except that of $\delta g_{\mathbf{v}\mathbf{v}}^{(0)}$ vanish at \eqref{omega-critical}, hence the critical frequency remains unchanged compared with the case in the absence of the Cotton tensor contribution. At this critical frequency $\omega_\star$, the coefficient of $\delta g_{\mathbf{v}\mathbf{v}}^{(0)}$ in \eqref{Linearized-EoM} leads to the following constraint
	\begin{align}
		\left(\left(\alpha_\mu k-i(1-\nu_\mu)r_-\right)^2+r_+^2 (1+\sigma_\mu)^2\right) (-i \alpha_\mu  k-(1-\nu_\mu) r_-+\lambda r_+ (1+\sigma_\mu))=0
	\end{align}
	Interestingly, we find that the two critical values of $k$ given in \eqref{kstartwo} are still solutions of the above constraint equation. This behavior is physically reasonable, since these two critical values ensure that the perturbed background remains an AdS$_3$ geometry, rendering the variation of the Cotton tensor trivial. However, when the Cotton tensor is included, one additional solution emerges:
	\begin{align}\label{kstarm}
		k_\star=\frac{i (r_-(1-\nu_\mu) -\lambda\, r_+(1+\sigma_\mu))}{\alpha_\mu}=\frac{2\pi i }{\beta  \left(1-\Omega ^2\right)}\left(\Omega-\frac{\lambda-\frac{4 \pi ^2 \mu }{\beta _+ \beta _- \lambda }}{\sqrt{1-\frac{8 \pi ^2 \mu }{\beta _+ \beta _-}+\left(\frac{4\pi^2\mu}{\beta_+\beta_-\lambda}\right)^2}} \right)
	\end{align}
	From \eqref{Pole-skipping-point}, we directly read off the chaos parameters as follows
	\begin{align}
		\lambda_L=\frac{2\pi}{\beta}~~,~~\frac{\left(1-\Omega^2\right)}{v_B}=\Omega \pm \frac{1\pm \frac{4 \pi ^2 \mu }{\beta _+ \beta _- \lambda }}{\sqrt{1-\frac{8 \pi ^2 \mu }{\beta _+ \beta _-}+\left(\frac{4\pi^2\mu}{\beta_+\beta_-\lambda}\right)^2}}\,,\,{\Omega-\frac{\lambda-\frac{4 \pi ^2 \mu }{\beta _+ \beta _- \lambda }}{\sqrt{1-\frac{8 \pi ^2 \mu }{\beta _+ \beta _-}+\left(\frac{4\pi^2\mu}{\beta_+\beta_-\lambda}\right)^2}}}\label{parameters-comoving}
	\end{align}
	Therefore, we find that the Lyapunov exponent saturates the chaos bound in the infalling EF coordinates with the co-moving spatial coordinate $\phi$. The three values of the butterfly velocity correspond to the left-moving, right-moving and the massive modes respectively.	We may obtain the frequencies in the Schwarzschild coordinates from \eqref{frequency-transformations} as follows
	\begin{align}
		\omega_\star^\textrm{Sch}&=\frac{2 \pi i}{\beta(1-\Omega^2)}\left(1 \pm \Omega\frac{1\pm \frac{4 \pi ^2 \mu }{\beta _+ \beta _- \lambda }}{\sqrt{1-\frac{8 \pi ^2 \mu }{\beta _+ \beta _-}+\left(\frac{4\pi^2\mu}{\beta_+\beta_-\lambda}\right)^2}}\right)\,,
	\end{align}
	for the left and right moving massless modes, and
	\begin{align}
		\omega_\star^\textrm{Sch}&=\frac{2\pi i }{\beta  \left(1-\Omega ^2\right)}\left(1-\Omega\frac{\lambda-\frac{4 \pi ^2 \mu }{\beta _+ \beta _- \lambda }}{\sqrt{1-\frac{8 \pi ^2 \mu }{\beta _+ \beta _-}+\left(\frac{4\pi^2\mu}{\beta_+\beta_-\lambda}\right)^2}} \right)\,,
	\end{align}
	for the massive mode. In the non-anomalous limit $\lambda\to \infty$ both the frequency and wave-vectors diverge linearly with $\lambda$, so that the nontrivial contribution of the massive mode vanishes as expected. The Lyapunov exponents and the butterfly velocities in the Schwarzschild coordinates may now be read off from \eqref{Pole-skipping-point} as follows:
	\begin{align}
		\lambda_L^\pm=\frac{2 \pi}{\beta(1-\Omega^2)}\left(1 \pm \Omega\frac{1\pm \frac{4 \pi ^2 \mu }{\beta _+ \beta _- \lambda }}{\sqrt{1-\frac{8 \pi ^2 \mu }{\beta _+ \beta _-}+\left(\frac{4\pi^2\mu}{\beta_+\beta_-\lambda}\right)^2}}\right)\,,\,
		v_B^\pm=\frac{\Omega  \left(1\pm \frac{4 \pi ^2 \mu }{\beta _+ \beta _- \lambda }\right)\pm\sqrt{1-\frac{8 \pi ^2 \mu }{\beta _+ \beta _-}+\left(\frac{4\pi^2\mu}{\beta_+\beta_-\lambda}\right)^2}}{\left(1\pm \frac{4 \pi ^2 \mu }{\beta _+ \beta _- \lambda }\right)\pm \Omega  \sqrt{1-\frac{8 \pi ^2 \mu }{\beta _+ \beta _-}+\left(\frac{4\pi^2\mu}{\beta_+\beta_-\lambda}\right)^2}}\,,\label{chaos-parameters-Sch-pm}
	\end{align}
	for the left (right) moving modes, and
	\begin{align}
		\lambda_L^m&=\frac{2 \pi}{\beta(1-\Omega^2)}\left(1- \frac{\Omega  \left(\lambda -\frac{4 \pi ^2 \mu }{\beta _+ \beta _- \lambda }\right)}{\sqrt{1-\frac{8 \pi ^2 \mu }{\beta _+ \beta _-}+\left(\frac{4\pi^2\mu}{\beta_+\beta_-\lambda}\right)^2}}\right)\,,\,
		v_B^m=\frac{\Omega  \left(\lambda -\frac{4 \pi ^2 \mu }{\beta _+ \beta _- \lambda }\right)-\sqrt{1-\frac{8 \pi ^2 \mu }{\beta _+ \beta _-}+\left(\frac{4 \pi ^2 \mu }{\beta _+ \beta _- \lambda }\right)^2}}{\lambda-\frac{4 \pi ^2 \mu }{\beta _+ \beta _- \lambda }-\Omega  \sqrt{1-\frac{8 \pi ^2 \mu }{\beta _+ \beta _-}+\left(\frac{4 \pi ^2 \mu }{\beta _+ \beta _- \lambda }\right)^2}}\,,\label{chaos-parameters-Sch-m}
	\end{align}
	for the massive mode.

	\subsubsection{Hagedorn behavior and various limits}
	
	\paragraph{Hagedorn behavior:} Interestingly, there exists a range of of $\mu$ values for which all the Lyapunov exponents and butterfly velocities become complex-valued. Note that the general expressions in \eqref{chaos-parameters-Sch-pm} and \eqref{chaos-parameters-Sch-m} exhibit a common square-root discriminant,
	\begin{align}\label{squareroot}
		\Delta(\mu)\equiv
		1-\frac{8\pi^2\mu}{\beta_+\beta_-}
		+\frac{16\pi^4\mu^2}{\beta_+^2\beta_-^2\lambda^2}\,.
	\end{align}
	The reality of the Lyapunov exponents and butterfly velocities is governed by the sign of
	$\Delta(\mu)$. The zeroes of this discriminant are given by
	\begin{align}
		\mu_{1,2}=
		\frac{\beta_+\beta_-}{4\pi^2}\,
		\lambda\left(\lambda\mp \sqrt{\lambda^2-1}\right),
		\label{Hagedorn}
	\end{align}
	Remarkably, these critical values correspond to the generalized Hagedorn behaviour of the deformed anomalous CFT$_2$, as pointed out in \cite{Basu:2025fsf}. For $\lambda>1$, one has
	$0<\mu_1<\mu_2$, whereas for $\lambda<-1$ the ordering is reversed,
	$0<\mu_2<\mu_1$. Therefore, in both cases, the interval between $\mu_1$ and $\mu_2$
	does not admit real-valued chaos parameters: the Lyapunov exponents and butterfly velocities become complex within this interval, indicating a genuine Hagedorn-type obstruction. At the critical points \(\mu=\mu_{1,2}\), the square root vanishes, and all three butterfly velocities coalesce to the angular velocity of the background,
	\begin{equation}
		v_B^{\pm,m}=\Omega\,.\label{vB-at-critical-mu}
	\end{equation}
	This degeneracy is the chaotic analogue of the Hagedorn singularity in the energy
	spectrum: the effective butterfly cone collapses precisely at the same deformation values where the deformed thermodynamics system no longer admits a real continuation. In this sense, the Hagedorn scale provides the natural upper bound for the real-time chaotic response. We will comment further on this behavior in Section \ref{sec:shockwave-OTOC}. It is instructive to examine several limiting cases:
	\paragraph{Undeformed limit:} In the limit $\mu\to 0$, we recover the undeformed results reported in \cite{Liu:2020yaf}
	\begin{align}
		\lambda^\pm_L&=\frac{2\pi}{\beta_\mp}~~,~~v^\pm_B=\pm 1\,,\notag\\
		\lambda^m_L&=\frac{2\pi\left(1-\lambda\,\Omega\right)}{\beta(1-\Omega^2)}~~,~~v^m_B=\frac{1-\lambda\,\Omega}{\Omega-\lambda}\,.
	\end{align}
	This provides a strong consistency check of our analysis.
	\paragraph{Non-anomalous limit:} Furthermore, in the non-anomalous limit $\lambda\to\infty$, the massive branch decouples, and we recover the chaos parameters reported in \cite{Basu:2025exh}
	%	\begin{align}
		%		\lambda_L^\pm=\frac{2\pi}{\beta(1-\Omega^2)}\left(1\mp\frac{\Omega}{\sqrt{1-\frac{8\pi^2\mu}{\beta_+\beta_-}}}\right)~~,~~v_B^\pm=\frac{1+\Omega\sqrt{1-\frac{8\pi^2\mu}{\beta_+\beta_-}}}{\Omega\pm\sqrt{1-\frac{8\pi^2\mu}{\beta_+\beta_-}}}
		%	\end{align}
	%	\textcolor{red}{corrected:}
	\begin{equation}
		\lambda_L^\pm=\frac{2\pi}{\beta\left(
			1-\Omega^2\right)} \left(
		1\pm\frac{\Omega}{\sqrt{1-\frac{8\pi^2\mu}{\beta_+\beta_-}}}\right),\quad v_B^\pm=\frac{\Omega\pm\sqrt{1-\frac{8\pi^2\mu}{\beta_+\beta_-}}}{1\pm\Omega\sqrt{1-\frac{8\pi^2\mu}{\beta_+\beta_-}}},
	\end{equation}
	for the left and right moving modes. In particular, the
	massless modes reproduce the familiar chiral asymmetry induced by rotation, while the
	massive mode ceases to represent an independent propagating channel.
	\paragraph{Chiral limit:} In the chiral limits $\lambda\to \pm 1$, we have 
	\begin{align}
		\begin{cases}
			\lambda_L^+=\frac{2\pi}{\beta_-}~~,~~v_B^+=1~~&\textrm{for}~\lambda\to-1\notag\\
			\lambda_L^-=\frac{2\pi}{\beta_+}~~,~~v_B^-=-1~~&\textrm{for}~\lambda\to+1
		\end{cases}
	\end{align}
	whereas
	\begin{align}
		\lambda_L^m=\frac{2\pi}{\beta_\pm}~~,~~v^m_B=\pm 1~~\textrm{as}~\lambda\to\pm 1
	\end{align}
	Therefore, in the chiral limits, the massive mode degenerates into either the left- or right-moving mode. Moreover, the chaos parameters become entirely independent of the $\TTbar$ deformation parameter $\mu$. This behavior is physically expected: in the chiral limits, only a single chiral component of the stress tensor survives, implying that the $\TTbar$ operator becomes null.
	
	\subsubsection{Non-rotating ensemble}\label{sec:non-rotating}
	
	For simplicity, we will analyze the behavior of the chaos parameters in case of the non-rotating ensemble. For the non-rotating black hole with $\Omega=0$, we find that all the Lyapunov exponents corresponding to the massless and massive modes saturate the MSS bound \cite{Maldacena:2015waa}, namely $\lambda_L=\frac{2\pi}{\beta}$. On the other hand, the butterfly velocities in the non-rotating case are given by
	\begin{align}
		v_B^{\pm}=\pm\frac{\sqrt{1-2\hat\mu+\hat\mu^2/\lambda^2}}{1\pm \hat\mu/\lambda}~~,~~
		v_B^{m}=-\frac{\sqrt{1-2\hat\mu+\hat\mu^2/\lambda^2}}{\lambda-\hat\mu/\lambda}\,,\label{vB-non-rotating}
	\end{align}
	where we have introduced the dimensionless deformation parameter
	\begin{equation}
		\hat\mu \equiv \frac{4\pi^2\mu}{\beta^2}\,.
	\end{equation}
	Notice that, due to the presence of gravitational anomaly, the maginitudes of butterfly velocities of the left and right moving modes are unequal even in the absence of rotation\footnote{Note that in the absence of the $\TTbar$ deformation, the maginitudes of butterfly velocities of the left and right moving modes are equal \cite{Liu:2020yaf}.}. We assume throughout that $|\lambda|>1$, consistent with the parameter regime considered in this work. The common square-root factor \eqref{squareroot} factorizes as
	\begin{equation}
		1-2\hat\mu+\frac{\hat\mu^2}{\lambda^2}
		=\frac{(\hat\mu-\hat\mu_-)(\hat\mu-\hat\mu_+)}{\lambda^2}\,,
		\qquad
		\hat\mu_\pm=\lambda^2\pm |\lambda|\sqrt{\lambda^2-1}\,.
		\label{eq:branchpoints}
	\end{equation}
	\begin{figure}
		\centering
		\includegraphics[width=0.75\linewidth]{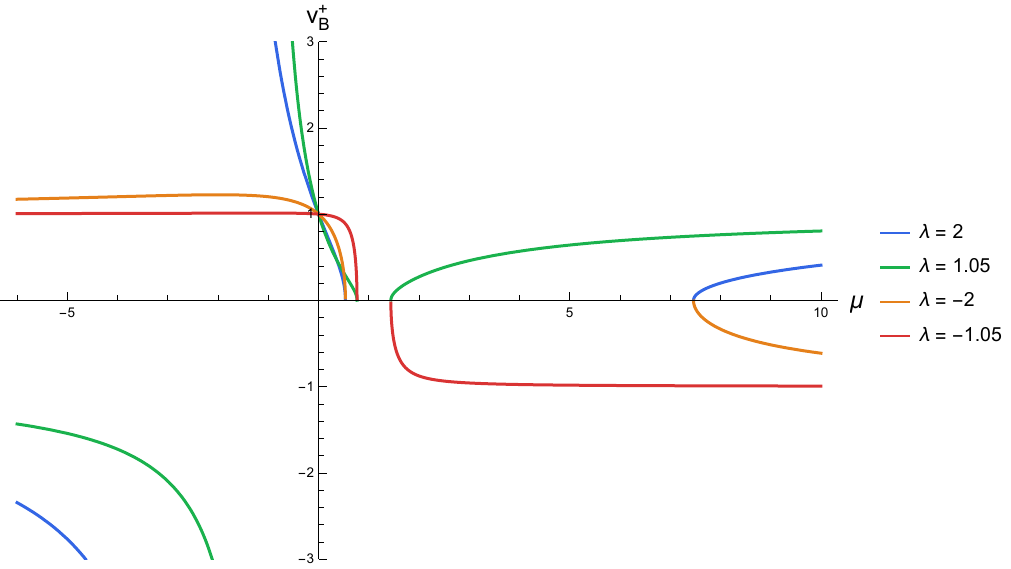}
		\includegraphics[width=0.75\linewidth]{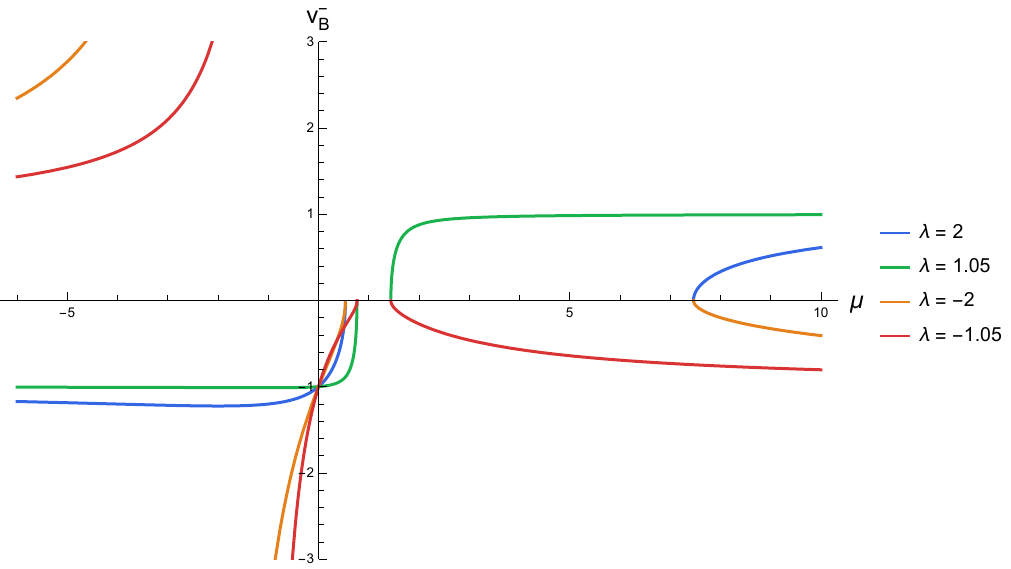}
		\caption{Butterfly velocities for the left- and right-moving modes in the non-rotating black hole. We have set $\beta=2\pi$.}
		\label{fig:nr-vbp}
	\end{figure}
	\begin{figure}[ht]
		\centering
		\includegraphics[width=0.75\linewidth]{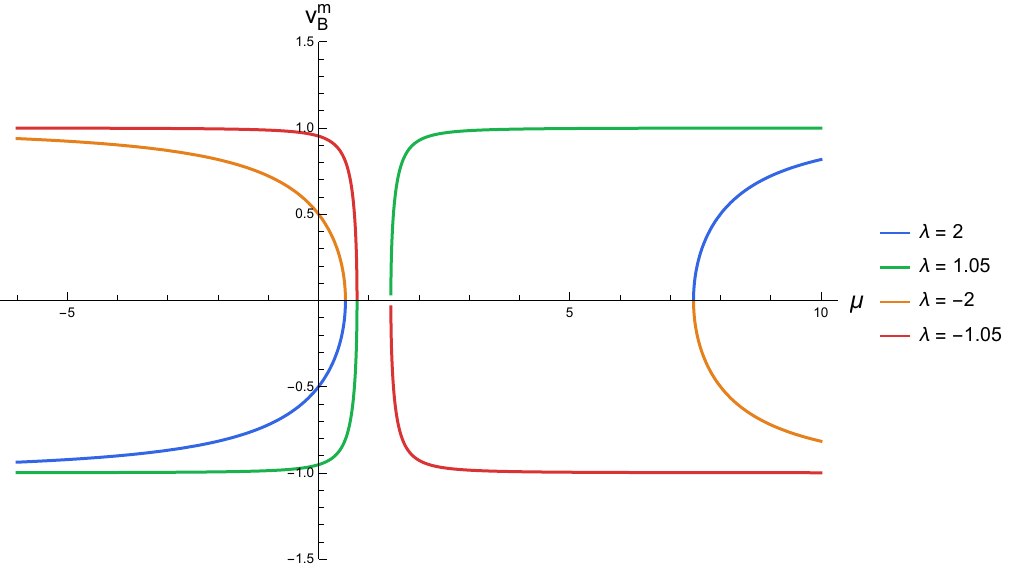}
		\caption{Butterfly velocity for the massive mode in the non-rotating black hole. We have set $\beta=2\pi$.}
		\label{fig:nr-vbm}
	\end{figure}
	Therefore the velocities are real only on the two disconnected sheets
	\begin{equation}
		\hat\mu\le \hat\mu_- \qquad \text{or} \qquad \hat\mu\ge \hat\mu_+\,,
	\end{equation}
	while the interval $\hat\mu_-\!<\hat\mu<\hat\mu_+$ is physically forbidden. This is nothing but the non-spinning analogues of the Hagedorn window \eqref{Hagedorn}. Since $\hat\mu=0$ lies on the lower sheet, the physical continuation from the undeformed theory terminates at the lower branch point $\hat\mu=\hat\mu_-$. The upper sheet is disconnected from the undeformed regime and should not be interpreted as a smoothly reachable second phase. In particular, the vanishing of the common square root at $\hat\mu=\hat\mu_\pm$ implies that all three butterfly velocities collapse to zero at the branch points, so the effective butterfly cone closes there.
	
	It is also instructive to investigate the pole structure of the butterfly velocities. The denominators in \eqref{vB-non-rotating} vanish at
	\begin{equation}
		\hat\mu=\mp\lambda \quad (\textrm{for~~}v_B^\pm)~~,~~
		\hat\mu=\lambda^2 \quad (\textrm{for~~}v_B^m)\,.
	\end{equation}
	These poles are not all accessible on the physical real branch. For $\lambda>1$, the pole of $v_B^+$ lies on the accessible real axis, while the poles of $v_B^-$ and $v_B^m$ fall inside the forbidden interval $\left(\hat{\mu}_-,\hat{\mu}_+\right)$. For $\lambda<-1$, the roles of $v_B^+$ and $v_B^-$ are interchanged, see Fig.\ref{fig:nr-vbp} for illustration. The massive-mode pole at $\hat\mu=\lambda^2$ always remains hidden inside the forbidden gap, so $v_B^m$ never develops a real divergence on the physical sheet, see Fig.\ref{fig:nr-vbm} for illustration
	
	%The massive branch is not a simple rescaling of the left-moving one. Instead,
	%\begin{equation}
	%v_B^-=\frac{\lambda^2-x}{\lambda-x}\,v_B^m\,,
	%\end{equation}
	%so the two branches have distinct analytic structures even though they share the same square-root factor. This distinction is physically important: the anomaly splits the propagation channels, but the deformation scale is still governed by the same universal branch points.
	
	Finally, we analyze potential deviations from the Mezei-Stanford bound \cite{Mezei:2016wfz}, which is governed by the sign of $\hat\mu$. One finds
	\begin{equation}
		(v_B^\pm)^2
		=1-\frac{2\hat\mu\left(1\pm\lambda^{-1}\right)}{(1+\hat\mu/\lambda)^2}\,
		~~,~~
		(v_B^m)^2
		=1-\frac{\lambda^2(\lambda^2-1)}{(\lambda^2-\hat\mu)^2}\,.
	\end{equation}
	Since $|\lambda|>1$ implies $1\pm \lambda^{-1}>0$, the two chiral butterfly velocities are superluminal for $\hat\mu<0$ and subluminal for $\hat\mu>0$, within their real parameter domain. In contrast, the massive mode remains always subluminal on the real domain. Therefore, the anomaly induces a genuine chiral asymmetry: the two light-cone branches react oppositely to the sign of the deformation parameter, while the massive branch remains universally bounded by unity. The superluminal phenomenon of chaotic propagation can be attributed to the nonlocal structure of the  $\TTbar$ deformed theory, a behavior previously reported in \cite{Basu:2025exh}. As discussed in \cite{Cardy:2018sdv,McGough:2016lol}, the signal propagation speed in $\TTbar$ deformed theories can become superluminal for $\mu<0$, which extends naturally to the present anomalous setting under consideration. Similar to the discussion in \cite{Basu:2025exh}, such seemingly pathological features may be interpreted through by mapping the $\TTbar$ deformed CFT to an undeformed CFT placed on a curved spacetime (See Section.\ref{sec:CFT-analysis}) with modified causal structure, leading to the notions of effective butterfly cones and renormalized butterfly velocities. At $\hat\mu=0$ one recovers
	\begin{equation}
		v_B^\pm(0)=\pm 1~~,~~v_B^m(0)=-\frac{1}{\lambda}\,,
	\end{equation}
	so the undeformed limit reproduces the relativistic light cone for the chiral sectors. The deformation then shifts the two chiral branches apart in an asymmetric manner, directly manifesting the effect of the gravitational anomaly. The main qualitative conclusions are summarized as follows:
	\begin{itemize}
		\item The non-rotating limit preserves the saturation of the chaos bound for the Lyapunov exponent. 
		\item The combined effect of $\TTbar$ deformation and gravitational anomaly splits the two chiral butterfly velocities even in the absence of rotation ($\Omega=0$).
		\item The Hagedorn branch points $\hat{\mu}_\pm$	delineate the parameter domain where real-time chaotic quantities are well-defined. 
	\end{itemize}
	These features persist in the subsequent shockwave analysis, where the same discriminant and ploe structures govern the behaviour of the transverse OTOC profile.

	\subsection{Shockwaves and OTOC} \label{sec:shockwave-OTOC}
	
	In this subsection, we briefly perform the shockwave analysis for the $\TTbar$ deformed BTZ black hole \eqref{deformed metric} in the presence of the gravitational Chern-Simons term. We first construct the Kruskal extension of the deformed rotating black hole metric \eqref{Co-moving-metric}. Introducing the in-going and out-going coordinates $(u,v)=(t-r_\star,t+r_\star)$, where $r_\star$ denotes the tortoise coordinate \eqref{rstar}, the metric takes the form,
	\begin{align}
		\d s^2=-\frac{r_h^2(r^2-r_+^2)}{r_+^2(1+\sigma_\mu)^2}\d u\d v&+\frac{r_-\left(r^2-r_+^2\right)\left(1-\nu_\mu\right)}{r_+\alpha_\mu\left(1+\sigma_\mu\right)}\left(\d u+\d v\right)\,\d\phi\notag\\
		&+\frac{1}{r_h^2\alpha_\mu^2}\left(r_+^2\left(1+\sigma_\mu\right)^2\left(r^2-r_-^2\right)-r_-^2\left(1-\nu_\mu\right)^2\left(r^2-r_+^2\right)\right)\d\phi^2\,.\label{in-out-metric}
	\end{align}
	The surface gravity at the (outer) horizon can be obtained as follows
	\begin{align}
		\kappa_s^2=-\frac{1}{2}\left|\left(\nabla^\mu\chi^\nu\right)\left(\nabla_\mu\chi_\nu\right)\right|_{r=r_+}\Rightarrow \kappa_s=\frac{r_h^2}{r_+(1+\sigma_{\mu })}=\frac{2\pi}{\beta}\,,
	\end{align}
	where $\chi=\partial_u$, and in the last equality, we have used \eqref{beta-Omega}.
	We then set up affine coordinates $(U,V)$ at the outer horizon,
	\begin{align}\label{UV}
		U=-e^{-\kappa_s u}~~,~~V=e^{\kappa_s v}\,.
	\end{align}
	These coordinates cover the right exterior $U<0, V>0$. The left exterior is obtained by flipping the signs of $(U,V)$. The explicit coordinate transformations from the Schwarzschild coordinates read,
	\begin{equation}
		U=-\left(\frac{\sqrt{r^2-r_-^2}-r_h}{\sqrt{r^2-r_-^2}+r_h}\right)^{1/2}e^{-\kappa_s t},\quad V=\left(\frac{\sqrt{r^2-r_-^2}-r_h}{\sqrt{r^2-r_-^2}+r_h}\right)^{1/2}e^{\kappa_s t}\,,
	\end{equation}
	or equivalently,
	\begin{equation}
		r=\frac{\sqrt{r_h^2\left(1-U V\right)^2+r_-^2\left(1+UV\right)^2}}{1+UV},\quad t=\frac{1}{2\kappa_s}\log\left(-\frac{V}{U}\right)
	\end{equation}
    In the Kruskal coordinate system, the deformed BTZ metric \eqref{Co-moving-metric} takes the form
	\begin{align}
		\d s^2=-\frac{4\d U\,\d V}{(1+U V)^2}&+\frac{4r_hr_-(1-\nu_\mu)(V\d U-U\d V)\d\phi}{\alpha_\mu (1+UV)^2}\notag\\&+\frac{r_h^2\left[4r_-^2(1-\nu_\mu)^2UV+r_+^2(1+\sigma_\mu)^2(1-UV)^2\right]}{\alpha_\mu^2(1+UV)^2}\d\phi^2
	\end{align}
	We now perturb the black hole by inserting a localized shockwave with stress-energy tensor
	\begin{align}
		T^\textrm{shock}_{UU}=E_0e^{\kappa_s t}\delta(U)\delta(\phi)\,,
	\end{align}
	where $E_0$ is the asymptotic energy of the shockwave, and the factor $e^{\kappa_s t}$ represents the large blue-shift experienced by the infalling matter. As a result, the perturbed spacetime has the following metric \cite{Dray:1984ha,Sfetsos:1994xa}
	\begin{align}
		\d s^2\to \d s^2+\frac{4}{(1+U V)^2}\delta(U)h(\phi)\d U^2\,,
	\end{align}
	where $h(\phi)$, referred to as the shockwave profile, encodes the backreaction of the shockwave on the background geometry and characterizes the horizon shift induced by the shockwave insertion. Then the $UU$-component of the TMG equation of motion leads to the following differential equation,
	\begin{align}
		&-\alpha_\mu^3 r_h^3 h'''\left(\phi\right)+\alpha_\mu r_h^2 \left(r_h\left(-3r_-^2\left(1-\nu_\mu\right)^2+2r_-r_+\lambda\left(1-\nu_\mu\right)\left(1+\sigma_\mu\right)+r_+^2\left(1+\sigma_\mu\right)^2\right)h'\left(\phi\right)\right.\notag\\
		&\left.-\alpha_\mu r_h \left(3r_-\left(-1+\nu_\mu\right)+r_+\lambda\left(1+\sigma_\mu\right)\right)h''\left(\phi\right)\right)\notag\\
		&+r_h^3\left(-r_-^3\left(-1+\nu_\mu\right)^3-r_-^2r_+\lambda\left(1-\nu_\mu\right)^2\left(1+\sigma_\mu\right)+r_-r_+^2\left(-1+\nu_\mu\right)\left(1+\sigma_\mu\right)^2+r_+^3\lambda\left(1+\sigma_\mu\right)^3\right)h\left(\phi\right) \notag\\
		&=4\pi G E_0 e^{\kappa t}r_+^3r_h^3\lambda\left(1+\sigma_\mu\right)^3\delta(\phi).\label{DE}
	\end{align}
    The general solution is given by,
	\begin{align}
		h(\phi)=c_1\,e^{\frac{r_-(1-\nu_\mu)+r_+(1+\sigma_\mu)}{\alpha_\mu}\phi}+c_2\,e^{\frac{r_-(1-\nu_\mu)-r_+(1+\sigma_\mu)}{\alpha_\mu}\phi}+c_3\,e^{\frac{r_-(1-\nu_\mu)-\lambda r_+(1+\sigma_\mu)}{\alpha_\mu}\phi},
	\end{align}
	where the three coefficients $c_i$ are constants ton be determined from the boundary condition and the compactness of the spatial coordinate. Note that we adopt a non-compact spatial coordinate here. In order to obtain the correct solution, we  impose the following falloff boundary condition
	\begin{equation}
		\lim_{|\phi|\rightarrow \infty}h\left(\phi\right)\rightarrow 0\,.
	\end{equation}
	In this case, there are three possible solutions,
	\begin{itemize}
		\item[$\bullet$] Solution \uppercase\expandafter{\romannumeral1}:  \begin{equation}
			h\left(\phi\right)=\begin{cases}
				\left(\lambda -1\right) e^{k_+\phi}, & \phi<0,\\
				\left(\lambda+1\right)e^{k_-\phi}-2 e^{k_m\phi}, & \phi>0.\\
			\end{cases}
		\end{equation}
		\item[$\bullet$] Solution \uppercase\expandafter{\romannumeral2}:  \begin{equation}
			h\left(\phi\right)=\begin{cases}
				\left(\lambda -1\right) e^{k_+\phi}-\left(\lambda +1\right) e^{k_-\phi}, & \phi<0,\\
				-2 e^{k_m\phi}, & \phi>0.\\
			\end{cases}
		\end{equation}
		\item [$\bullet$]Solution \uppercase\expandafter{\romannumeral3}: \begin{equation}
			h\left(\phi\right)= \begin{cases}
				\left(\lambda -1\right) e^{k_+\phi}+2 e^{k_m\phi}, & \phi<0,\\
				\left(\lambda +1\right) e^{k_-\phi}, & \phi>0.\\
			\end{cases}
		\end{equation} 
	\end{itemize}
	where $k_\pm,k_m$ are given by the absolute values of the pole-skipping momenta \eqref{kstartwo} and  \eqref{kstarm}, respectively, i.e.,
	\begin{equation}
		\begin{aligned}
			k_\pm=&\frac{2\pi}{\beta\left(1-\Omega^2\right)}\left(\Omega\pm \frac{1\pm \frac{4\pi^2\mu}{\beta_+\beta_- \lambda}}{\sqrt{1-\frac{8\pi^2 \mu}{\beta_+\beta_-}+\left(\frac{4\pi^2 \mu}{\beta_+\beta_- \lambda}\right)^2}}\right),\\
			k_m=&\frac{2\pi}{\beta\left(1-\Omega^2\right)}\left(\Omega- \frac{\lambda - \frac{4\pi^2\mu}{\beta_+\beta_- \lambda}}{\sqrt{1-\frac{8\pi^2 \mu}{\beta_+\beta_-}+\left(\frac{4\pi^2 \mu}{\beta_+\beta_- \lambda}\right)^2}}\right),
		\end{aligned}
	\end{equation}
	and we have rescaled the function $h\left(\phi\right)$ by an overall constant factor via,
	\begin{equation}
		h\left(\phi\right)\rightarrow \frac{2\pi G r_+ \lambda\left(1+\sigma_\mu\right)}{\alpha_\mu \left(\lambda^2-1\right)}E_0 e^{\kappa_s t} h\left(\phi\right).
	\end{equation}
	\begin{table}[ht]
		\centering
		\begin{tabular}{|c|c|c|c|c|c|}
			\hline
			$\lambda$& $\mu$ & $k_+$ & $k_-$ & $k_m$ & Solution \\
			\hline
			\multirow{3}{*}{$\lambda<-1$} & $\mu\leq\mu_{c,2}$ & $>0$ & $\geq 0$ & $>0$ &$\times$  \\
			\cline{2-6}
			~ & $\mu_{c,2}<\mu<\mu_2$ & $>0$ & $<0$ & $>0$ & \uppercase\expandafter{\romannumeral3}  \\
			\cline{2-6}
			~ & $\mu>\mu_{1}$ & $<0$ & $<0$ & $<0$ & $\times$ \\
			\hline
			\multirow{3}{*}{$1<\lambda<\lambda_c$} & $\mu\leq \mu_{c,1}$ & $\leq 0$ & $<0$ & $<0$ & $\times$ \\
			\cline{2-6}
			~ & $\mu_{c,1}<\mu<\mu_{1}$ & $>0$ & $<0$ & $<0$ & \uppercase\expandafter{\romannumeral1} \\
			\cline{2-6}
			~ & $\mu>\mu_{2}$ & $>0$ & $>0$ & $>0$ & $\times$ \\
			\hline
			\multirow{4}{*}{$\lambda=\lambda_c$} & $\mu\leq \mu_{c,1}$ & $\leq 0$ & $<0$ & $<0$ & $\times$ \\
			\cline{2-6}
			~ & \makecell[c]{$\mu_{c,1}<\mu<\mu_{c,2}$ \\ or $\mu_{c,2}<\mu<\mu_1$} & $>0$ & $<0$ & $<0$ & \uppercase\expandafter{\romannumeral1} \\
			\cline{2-6}
			~ & $\mu=\mu_{c,2}$ & $>0$ & $=0$ & $<0$ & $\times$ \\
			\cline{2-6}
			~ & $\mu>\mu_2$ & $>0$ & $>0$ & $>0$ & $\times$ \\
			\hline
			\multirow{5}{*}{$\lambda>\lambda_c$} & $\mu\leq \mu_{c,1}$ & $\leq 0$ & $<0$ & $<0$ & $\times$ \\
			\cline{2-6}
			~ &\makecell[c]{$\mu_{c,1}<\mu<\mu_{c,2}$ \\ or $\mu_{c,3}<\mu<\mu_1$}& $>0$ & $<0$ & $<0$ & \uppercase\expandafter{\romannumeral1}  \\
			\cline{2-6}
			~ & $\mu=\mu_{c,2}$ or $\mu=\mu_{c,3}$ & $>0$ & $=0$ & $<0$ & $\times$ \\
			\cline{2-6}
			~ & $\mu_{c,2}<\mu<\mu_{c,3}$ & $>0$ & $>0$ & $<0$ & \uppercase\expandafter{\romannumeral2} \\
			\cline{2-6}
			~ & $\mu>\mu_2$ & $>0$ & $>0$ & $>0$ & $\times$ \\
			\hline
		\end{tabular}
		\caption{Parameter ranges corresponding to solutions for $h(\phi)$ for a non-compact $\phi$.}
		\label{Table-1}
	\end{table}
	
\begin{figure}[ht]
		\centering
		\includegraphics[width=0.45\linewidth]{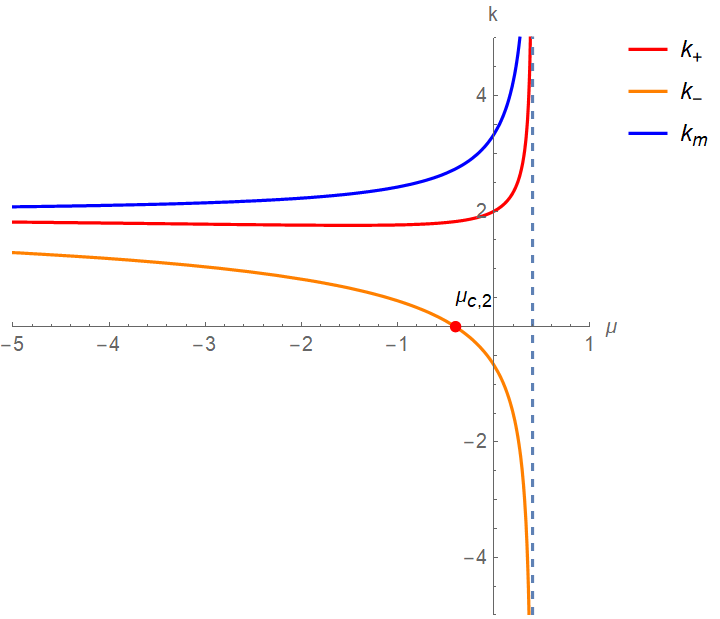}
		\includegraphics[width=0.45\linewidth]{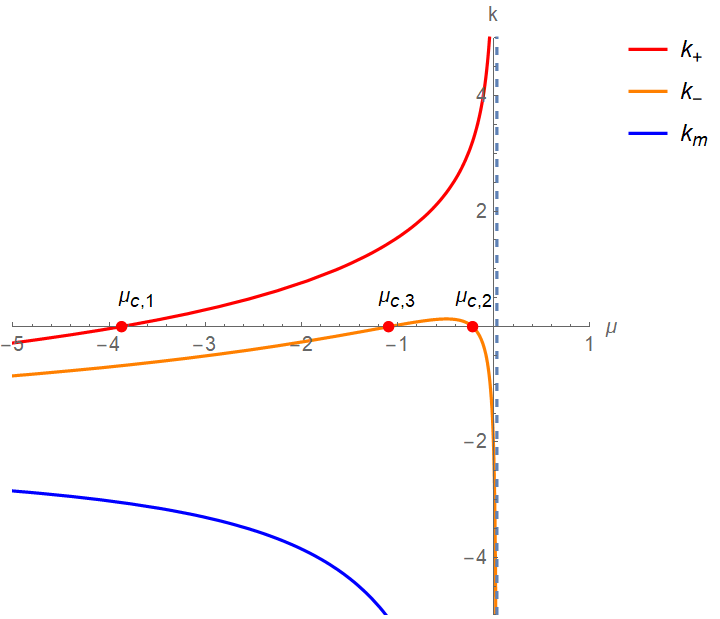}
		\caption{The illustration of Table \ref{Table-1}. We have set $\beta=2\pi,\Omega=1/2,\lambda=-2 ~\textrm{(left panel)}\,,8 ~\textrm{(right panel)}$. The two dashed lines are $\mu=\mu_{1}$ and $\mu=\mu_{2}$, respectively.}
		\label{fig:example1}
	\end{figure}

		The specific form of the solution depends on the value of the deformation parameter $\mu$, which is delineated in Table \ref{Table-1}. In this table, $\mu_1$ and $\mu_2$ are given by  \eqref{Hagedorn}, i.e. the zero points of the discriminant $\Delta\left(\mu\right)$ in \eqref{squareroot}.
	The value $\mu_{c,1}$ is the zero point of $k_+$, i.e.
	\begin{equation}
		\begin{aligned}
			\mu_{c,1}=-\frac{\beta^2 \lambda }{4\pi^2}\left(1+\lambda \Omega^2+\Omega\sqrt{\left(1+\lambda\right)\left(2+\left(\lambda-1\right)\Omega^2\right)}\right),
		\end{aligned}
	\end{equation}
	while $\mu_{c,2}$ and $\mu_{c,3}$ are the zero points of $k_-$, i.e.
	\begin{equation}
		\begin{aligned}
			\mu_{c,2}=&\:\frac{\beta^2 \lambda }{4\pi^2}\left(1-\lambda \Omega^2-\Omega\sqrt{\left(\lambda-1\right)\left(-2+\left(\lambda+1\right)\Omega^2\right)}\right),
			\\
			\mu_{c,3}=&\:\frac{\beta^2 \lambda }{4\pi^2}\left(1-\lambda \Omega^2+\Omega\sqrt{\left(\lambda-1\right)\left(-2+\left(\lambda+1\right)\Omega^2\right)}\right).
		\end{aligned}
	\end{equation}
	Moreover, $\lambda_c$ denotes the phase transition point associated with $\mu_{c,2}$ and $\mu_{c,3}$,
	\begin{equation}
		\lambda_{c}=-1+\frac{2}{\Omega^2}.\quad \mu_{c,2},\:\mu_{c,3}\notin\text{Reals, when } 1<\lambda<\lambda_{c},\quad
		\mu_{c,2},\:\mu_{c,3}\in\text{Reals, when } \lambda\geq \lambda_{c}.
	\end{equation}
	The critical values $\mu_{c,i}$ coorespond to the pole structure of the butterfly velocities, whose non-spinning analogues were investigated in Section \ref{sec:non-rotating}.
	Interestingly, the reality condition for the holographic entanglement entropy of the anomalous CFT with $\TTbar$ deformation discussed in \cite{Basu:2025fsf} exactly corresponds the ranges of the deformation parameter $\mu$ associated with the Solution \uppercase\expandafter{\romannumeral1} and Solution \uppercase\expandafter{\romannumeral3}, where $k_+>0,k_-<0$. Figure \ref{fig:example1} provides an illustration of the above mentioned prperties for a specific set of parameter values. The OTOC may now be obtained from the solutions $h(\phi)$ through the following relation \cite{Jahnke:2019gxr} \footnote{Note that, our expression of $C(t,\phi)$ misses a factor of $e^{\kappa_st}$ in front of $h(\phi)$ as compared to \cite{Jahnke:2019gxr}, since in our convention $h(\phi)$ already carries the same prefactor.},
	\begin{align}
		C(t,x)=1-\epsilon^{}_{VW}h(\Omega t-x),\label{OTOC-from-h}
	\end{align}
	which is governed by three sets of exponents and the corresponding velocities, depending on the relative strngths of the deformation and gravitational anomaly. As expected, it is easy to verify that the shockwave analysis leads to the same set of chaos parameters \eqref{parameters-comoving} as those obtained from the pole skipping analysis. This confirms the applicablity of the arguments presented in \cite{Wang:2022mcq,Chua:2025vig} in the presence of the non-local $\TTbar$ deformation as well as the higher-derivative corrections leading to the gravitational anomaly.
	
	\paragraph{Remarks:}
	\begin{itemize}
		\item In the ranges corresponding to \uppercase\expandafter{\romannumeral1} and \uppercase\expandafter{\romannumeral3} (i.e. HEE$>0$), we have
		\begin{equation}
			\lambda_L^-< \frac{2\pi }{\beta},\quad \lambda_L^+> \frac{2\pi }{\beta},\quad \lambda_L^m< \frac{2\pi }{\beta},
		\end{equation}
		For Solution \uppercase\expandafter{\romannumeral2},
		\begin{equation}
			\lambda_L^-> \frac{2\pi }{\beta},\quad \lambda_L^+> \frac{2\pi }{\beta},\quad \lambda_L^m< \frac{2\pi }{\beta}.
		\end{equation}
		\item Furthermore, in the ranges corresponding to \uppercase\expandafter{\romannumeral1} and \uppercase\expandafter{\romannumeral3} (i.e. HEE$>0$), $v_B^m$ behaves well, i.e.
		\begin{equation}
			|v_m|<1.
		\end{equation}
		In particular, once $\mu>0$, all butterfly velocities $v_B^{\pm,m}$ are physically acceptable, satisfying
		\begin{equation}
			0<v_{B}^{-}<1,\quad 0>v_{B}^{+}>-1.
		\end{equation}
		\item However, for $\mu<0$, the butterfly velocities can become superluminal, consistent with the discussion in Section \ref{sec:non-rotating}..
	\end{itemize}
	%-------------------------------------------------------------------
	\subsubsection{Effect of periodicity}
	
\begin{figure}[ht]
		\centering
		\includegraphics[width=0.45\linewidth]{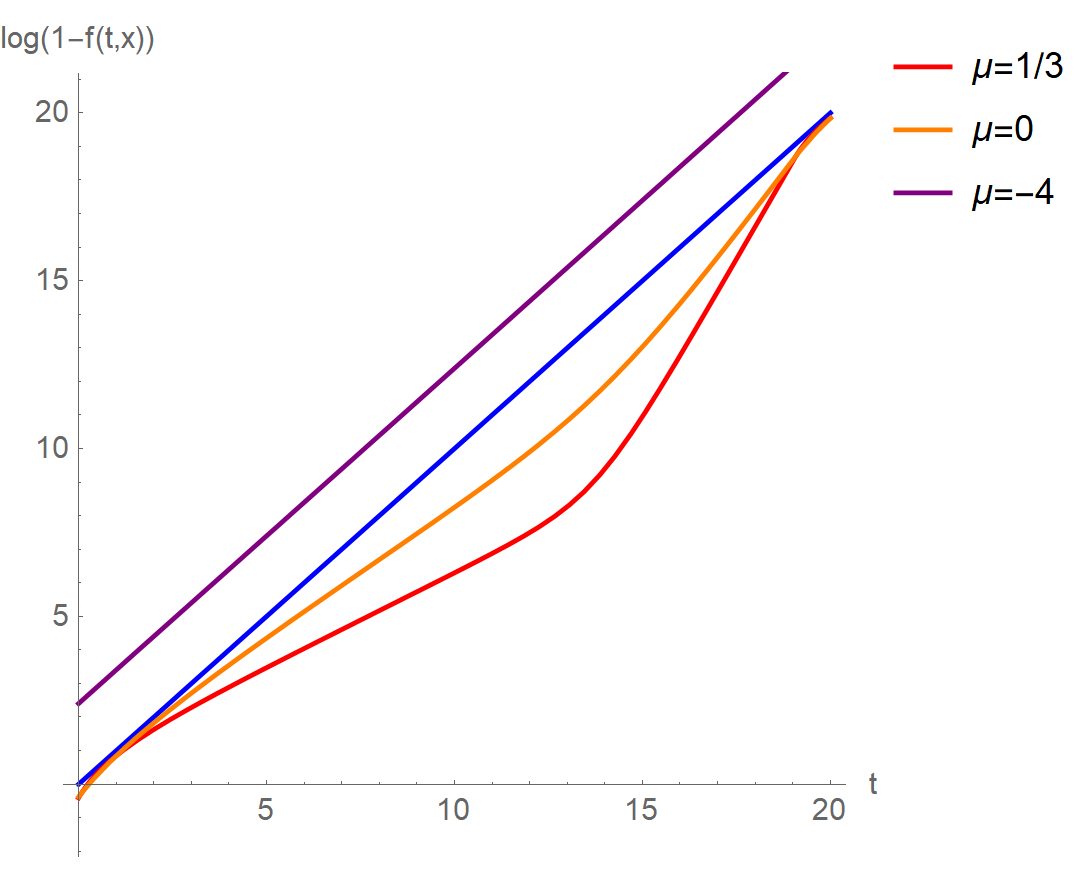}
		\caption{Modulation of the OTOC for different deformation parameters. we have set $\beta=2\pi,\Omega=0.33,\lambda=2$.}
		\label{fig:ffunction}
	\end{figure}

	As discussed in \cite{Jahnke:2019gxr,Mezei:2019dfv,Liu:2020yaf}, the above analysis remains valid only when the spatial direction is non-compact, which requires the high-temperature assumption. In contrast,  for intermediate temperatures, one should take into account the periodicity of the spatial coordinate to correctly extract the characteristic exponents governing the OTOC.
	Now we consider the periodicity of $\phi$, by replacing the delta function in \eqref{DE} with the periodic sum  $\sum_{n=0}^{\infty}\delta(\phi-2\pi n)$. We focus on the three types of solutions, depending on the value of the deformation parameter $\mu$:
	\begin{itemize}
		\item[$\bullet$] Solution \uppercase\expandafter{\romannumeral1}: 
		\begin{equation}
			\begin{aligned}
				h\left(\phi\right)=&\left(\lambda -1\right)\sum_{n=1}^{\infty}e^{k_+\left(\phi-2\pi n\right)}+\left(\lambda+1\right)\sum_{n=-\infty}^{0}e^{k_-\left(\phi-2\pi n\right)}-2\sum_{n=-\infty}^{0}e^{k_m\left(\phi-2\pi n\right)}\\
				=&\left(\lambda -1\right)\frac{e^{k_+\phi}}{e^{2\pi k_+}-1}-\left(\lambda+1\right)\frac{e^{k_- \phi}}{e^{2\pi k_-}-1}+2\frac{e^{k_m\phi}}{e^{2\pi k_m}-1}.
			\end{aligned}
		\end{equation}
		\item[$\bullet$] Solution \uppercase\expandafter{\romannumeral2}: 
		\begin{equation}
			\begin{aligned}
				h\left(\phi\right)=&\left(\lambda -1\right)\sum_{n=1}^{\infty}e^{k_+\left(\phi-2\pi n\right)}-\left(\lambda+1\right)\sum_{n=1}^{\infty}e^{k_-\left(\phi-2\pi n\right)}-2\sum_{n=-\infty}^{0}e^{k_m\left(\phi-2\pi n\right)}\\
				=&\left(\lambda -1\right)\frac{e^{k_+\phi}}{e^{2\pi k_+}-1}-\left(\lambda+1\right)\frac{e^{k_- \phi}}{e^{2\pi k_-}-1}+2\frac{e^{k_m\phi}}{e^{2\pi k_m}-1}.
			\end{aligned}
		\end{equation}
		\item[$\bullet$] Solution \uppercase\expandafter{\romannumeral3}:
		\begin{equation}
			\begin{aligned}
				h\left(\phi\right)=&\left(\lambda -1\right)\sum_{n=1}^{\infty}e^{k_+\left(\phi-2\pi n\right)}+2\sum_{n=1}^{\infty}e^{k_m\left(\phi-2\pi n\right)}+\left(\lambda+1\right)\sum_{n=-\infty}^{0}e^{k_-\left(\phi-2\pi n\right)}\\
				=&\left(\lambda -1\right)\frac{e^{k_+\phi}}{e^{2\pi k_+}-1}-\left(\lambda+1\right)\frac{e^{k_- \phi}}{e^{2\pi k_-}-1}+2\frac{e^{k_m\phi}}{e^{2\pi k_m}-1}.
			\end{aligned}
		\end{equation}
	\end{itemize}
	Therefore, there is a unique physical solution for the periodic spatial coordinate $\phi$. In Figure.\ref{fig:ffunction}, we analyze the behavior of the OTOC obtained from \eqref{OTOC-from-h} as a function of $t$
	with $x$ kept fixed. Similar to the undeformed case in \cite{Mezei:2019dfv}, the function $\log(1-\CC(t,x))$ grows on the average as $e^{2\pi t/\beta}$ with periodic modulation governed by the deformation parameter $\mu$ and Chern-Simons coupling $\lambda$. It is easy to verify that the average Lyapunov exponenet saturates the chaos bound \cite{Maldacena:2015waa}. Furthermore, the generalized chaos bound for rotating thermal ensembles proposed in \cite{Mezei:2019dfv} is also saturated,
	\begin{align}
		\left|\frac{(\del_t+\Omega \del_x)\CC(t,x)}{1-\CC(t,x)}\right|=\frac{2\pi}{\beta}\,,
	\end{align}
	even in  the presence of non-local $\TTbar$ deformation and the chiral effects of the gravitational anomaly.

	\subsubsection{High temperature limits}
	
	Now we study the high temperature behavior of the instantaneous Lyapunov exponent, which is defined as
	\begin{equation}
		\lambda_{\text{inst}}\left(t\right)=\frac{2\pi}{\beta}+\frac{\partial_t h\left(\Omega t\right)}{h\left(\Omega t\right)}.
	\end{equation}
	Recall that the temperature can not exceed the Hagedorn bound, otherwise the $\TTbar$ deformed energy spectrum will become complex and hence do not correspond to a physically relevant situation. Hence by high temperature we mean the Hagedorn temperature given by,
	\begin{equation}
		\beta\rightarrow \beta_{H}=\begin{cases}
			\frac{2 \pi  \mu }{\sqrt{\lambda  \left(\lambda -\sqrt{\lambda ^2-1}\right) \mu  \left(1-\Omega ^2\right)}}, & \text{when }\lambda>1,\\
			\frac{2 \pi  \mu }{\sqrt{\lambda  \left(\lambda +\sqrt{\lambda ^2-1}\right) \mu  \left(1-\Omega ^2\right)}}, & \text{when }\lambda<-1.\\
		\end{cases} 
	\end{equation}
	Note that only for Solution \uppercase\expandafter{\romannumeral1} and Solution \uppercase\expandafter{\romannumeral3} cases, the temperature can approach Hagedorn temperature. In the Hagedorn limit, the instantaneous Lyapunov exponent is given by,
	\begin{equation}
		\lambda_{\text{inst}}\left(t\right)=\begin{cases}
			\lambda_-, & \text{when }t\in[0,\frac{\pi\left(
				1+\lambda-\sqrt{\lambda^2-1}\right)}{\Omega } ),\\
			\lambda_+, & \text{when }t\in[\frac{\pi\left(
				1+\lambda-\sqrt{\lambda^2-1}\right)}{\Omega },\frac{2\pi}{\Omega} ),\\
		\end{cases}
	\end{equation}
	where $\lambda_\pm$ are given by \eqref{chaos-parameters-Sch-pm}. Furthermore, one can easily check that the averaged Lyapunov exponent satisfies
	\begin{equation}
		\langle \lambda_{\text{inst}}\left(t\right)\rangle=\frac{2\pi}{\beta},
	\end{equation}
	for any temperature. Remarkably, the contribution of the massive mode does not survive in the Hagedorn limit once the spatial coordinate is compactified.
	%%---------------------------------------------------
	%	\subsubsection{OTOC at the chiral point}
	\subsection{CFT analysis}\label{sec:CFT-analysis}
	
	In this subsection, we substantiate our holographic findings through a direct field theoretic analysis of the OTOC in the deformed anomalous field theory, following \cite{Perlmutter:2016pkf,Alishahiha:2016cjk}.
	As described in \cite{Basu:2025fsf}, we may rewrite the induced metric on the asymptotic boundary at $\tilde{\rho}=\tilde{\rho}_\infty$, in terms of the boundary conformal time as
	\begin{align}
		\d s^2_\textrm{bdy}=\frac{-\d t^2+\d\hat{\phi}^2}{\tilde\rho_\infty(1+2\tilde{\mu}\kappa\CL_\mu)(1+2\tilde{\mu}\bar\kappa\bar\CL_\mu)}:=\Omega_\omega^{-1}\Omega_{\bar\omega}^{-1}\frac{\d\omega\,\d\bar\omega}{\tilde\rho_\infty}
	\end{align}
	where we have switched to the Euclidean signature and defined the complex coordinate $\omega=\hat\phi+i\tau$, with the conformal spatial coordinate
	\begin{align}
		\hat{\phi}=\frac{(1+2\tilde{\mu}\kappa\CL_\mu)(1+2\tilde{\mu}\bar\kappa\bar\CL_\mu)}{(1-4\tilde{\mu}^2\kappa\bar\kappa\CL_\mu\bar\CL_\mu)}x+2\tilde{\mu}\frac{\kappa\CL_\mu-\bar\kappa\bar\CL_\mu}{1-4\tilde{\mu}^2\kappa\bar\kappa\CL_\mu\bar\CL_\mu}t\,.\label{conformal-frame}
	\end{align}
	Notably, in these coordinates the boundary metric is conformal to a \textit{twisted cylinder} $\CM$ \cite{Caputa:2013eka}, with constant conformal factors for the holomorphic and anti-holomorphic sectors
	\begin{align}
		\Omega_\omega=1+2\tilde{\mu}\kappa\CL_\mu~~,~~\Omega_{\bar\omega}=1+2\tilde{\mu}\bar\kappa\bar\CL_\mu
	\end{align}
	Therefore, we may interpret the holographic $\TTbar$-deformed anomalous thermal CFT as conformally equivalent to a undeformed CFT$_2$ defined on this twisted cylinder, with left and right moving central charges $c^{}_L$ and $c^{}_R$.
	In particular, the thermal identifications in these coordinates are given as
	\begin{align}
		(\omega,\bar{\omega})\sim(\omega+\beta_\omega,\bar{\omega}+\beta_{\bar\omega})
	\end{align}
	where 
	\begin{align}
		\beta_\omega=\frac{\pi}{\sqrt{\CL_\mu}}+2\pi\tilde{\mu}\kappa\sqrt{\CL_\mu}~~,~~\beta_{\bar\omega}=\frac{\pi}{\sqrt{\bar\CL_\mu}}+2\pi\tilde{\mu}\bar\kappa\sqrt{\bar\CL_\mu}
	\end{align}
	The temperature and the angular speed in this conformal frame are then given by\footnote{Note that the angular speed differs from the one in the deformed field theory in \eqref{beta-pm}, since it corresponds to the thermal identifications within the conformal frame \eqref{conformal-frame}. Nevertheless, the temperature remains the same, as the time coordinate is identical in both frames.}
	\begin{align}
		\beta&=\frac{\beta_\omega+\beta_{\bar\omega}}{2}=\frac{\pi}{2}\left(\frac{1}{\sqrt{\CL_\mu}}+\frac{1}{\sqrt{\bar\CL_\mu}}\right)+\pi\mu(\kappa\sqrt{\CL_\mu}+\bar\kappa\sqrt{\bar\CL_\mu})=\frac{2\pi}{r_h^2}r_+(1+\sigma_\mu)\,,\notag\\
		\hat\Omega&=\frac{\beta_\omega-\beta_{\bar\omega}}{\beta_\omega+\beta_{\bar\omega}}=\frac{\sqrt{\bar\CL_\mu}(1+2\mu\kappa\CL_\mu)-\sqrt{\CL_\mu}(1+2\mu\bar\kappa\bar\CL_\mu)}{\sqrt{\bar\CL_\mu}(1+2\mu\kappa\CL_\mu)+\sqrt{\CL_\mu}(1+2\mu\bar\kappa\bar\CL_\mu)}=\frac{r_-(1-\nu_\mu)}{r_+(1+\sigma_\mu)}
	\end{align}
	
	We may now map the twisted cylinder with complex coordinates $(\omega,\bar\omega)$ to the complex plane utilizing the conformal transformations
	\begin{align}
		z=e_{}^{\frac{2\pi \omega}{\beta_\omega}}~~,~~\bar z=e_{}^{\frac{2\pi\bar\omega}{\beta_{\bar\omega}}}\,.\label{CFT-map-to-cyl}
	\end{align}
	Recall that a faithful diagonostic of quantum chaos is given by the OTOC between pairs of Hermitian and unitary operators $W$ and $V$ \eqref{OTOC-defn} in the thermal state with inverse temperature $\beta$. Utlizing the transformations \eqref{CFT-map-to-cyl}, this four-point function may be mapped to the complex plane. Subsequently, the (normalized) four-point correlator on the complex plane may be evaluated in the large central charge limit (particularly, in the Regge limit or eikonal approximation \cite{Perlmutter:2016pkf}) using techniques developed in \cite{Fitzpatrick:2014vua}
	\begin{align}
		\CF(\eta,\bar\eta)&=\frac{\left<W(z_1,\bar z_1)V(z_2,\bar z_2)W(z_3,\bar z_3)V(z_4,\bar z_4)\right>}{\left<W(z_1,\bar z_1)W(z_3,\bar z_3)\right>\left<V(z_2,\bar z_2)V(z_4,\bar z_4)\right>}\notag\\&=2\pi i\sum_{\CO(\Delta,s)}\alpha_\CO^2\frac{\Gamma(\Delta,s)\Gamma(\Delta+s-1)}{\Gamma^4\left(\frac{\Delta+s}{2}\right)}\eta^{1-s}\left(\frac{\bar\eta}{\eta}\right)^{\frac{\Delta-s}{2}}
	\end{align}
	where the sum runs over conformal primary operators $\CO$, the cross-ratios $(\eta,\bar\eta)$ are given by
	\begin{align}
		\eta=\frac{z_{12}z_{34}}{z_{13}z_{24}}~~,~~\bar\eta=\frac{\bar z_{12}\bar z_{34}}{\bar z_{13}\bar z_{34}}\,,
	\end{align}
	and $\alpha_\CO$ is constituted out of the OPE coefficients in the CFT.
	Note that, in order to capture the correct time-ordering in \eqref{OTOC-defn}, it is required to introduce imfinitesimal Euclidean times $i\epsilon_k$ to each of the operators and make the choice $\epsilon_1<\epsilon_2<\epsilon_3<\epsilon_4$. Then, in the Regge regime, one finds \cite{Perlmutter:2016pkf,Craps:2021bmz}
	\begin{align}
		\eta=-e^{\frac{2\pi}{\beta_\omega}(\hat\phi-t)}\epsilon_{12}^\star\epsilon_{34}~~,~~\bar \eta=-e^{-\frac{2\pi}{\beta_{\bar\omega}}(\hat\phi+t)}\epsilon_{12}^\star\epsilon_{34}
	\end{align}
	For the case of a CFT$_2$ at finite temperature $\beta$ and angular speed $\hat\Omega$,
	\begin{align}
		\CF(\eta,\bar\eta)\approx2\pi i\sum_{\CO(\Delta,s)}\alpha_\CO^2\frac{\Gamma(\Delta,s)\Gamma(\Delta+s-1)}{\Gamma^4\left(\frac{\Delta+s}{2}\right)}\exp\left[\frac{2\pi\left(s-1+\hat\Omega(\Delta -1)\right) }{\beta(1-\hat\Omega ^2)}\left(t-\frac{\Delta-1 +(s-1) \hat\Omega}{s-1+(\Delta -1) \hat\Omega}\hat{\phi}\right)\right]\label{Regge-block-rotating}
	\end{align}
	One may now read off the Lyapunov exponent and the butterfly velocities in the conformal coordinates $(\hat{\phi},t)$ as follows
	\begin{align}
		\lambda^\textrm{conf}_L(\Delta,s)=\frac{2\pi\left(s-1+\hat\Omega(\Delta -1)\right) }{\beta(1-\hat\Omega ^2)}~~,~~v^\textrm{conf}_B(\Delta,s)=\frac{s-1+(\Delta -1) \hat\Omega}{\Delta-1 +(s-1) \hat\Omega}\,.
	\end{align}
	Transforming to the comoving coordinates using \cref{co-moving,conformal-frame}, we may find
	\begin{align}
		\CF(\eta,\bar\eta)\approx2\pi i\sum_{\CO(\Delta,s)}\alpha_\CO^2&\frac{\Gamma(\Delta,s)\Gamma(\Delta+s-1)}{\Gamma^4\left(\frac{\Delta+s}{2}\right)}\exp\Bigg[\frac{4 \sqrt{\CL_\mu\bar\CL_\mu} (s-1) }{\sqrt{\CL_\mu}(1+2\mu\bar\kappa\bar\CL_\mu)+\sqrt{\bar\CL_\mu}(1+2\mu\kappa\CL_\mu)}t\notag\\&-\frac{(\Delta-s)\sqrt{\CL_\mu}(1+2\mu\bar\kappa\bar\CL_\mu)+(\Delta+s-2)\sqrt{\bar\CL_\mu}(1+2\mu\kappa\CL_\mu)}{1-4\mu^2\kappa\bar\kappa\CL_\mu\bar\CL_\mu}\phi\Bigg]
	\end{align}
	from which we may obtain the chaos parameters in the co-moving frame as
	\begin{align}
		\lambda_L^\textrm{cm}&=\frac{4 \sqrt{\CL_\mu\bar\CL_\mu} (s-1) }{\sqrt{\CL_\mu}(1+2\mu\bar\kappa\bar\CL_\mu)+\sqrt{\bar\CL_\mu}(1+2\mu\kappa\CL_\mu)}=\frac{2\pi}{\beta}(s-1)\,,\notag\\
		v_B^\textrm{cm}&=\frac{4(s-1)\sqrt{\CL_\mu\bar\CL_\mu}(1-4\mu^2\kappa\bar\kappa\CL_\mu\bar\CL_\mu)}{(\sqrt{\CL_\mu}(1+2\mu\bar\kappa\bar\CL_\mu)+\sqrt{\bar\CL_\mu}(1+2\mu\kappa\CL_\mu))((\Delta-s)\sqrt{\CL_\mu}(1+2\mu\bar\kappa\bar\CL_\mu)+(\Delta+s-2)\sqrt{\bar\CL_\mu}(1+2\mu\kappa\CL_\mu))}\notag\\
		&=\frac{\frac{2\pi}{\beta}(s-1)}{(\Delta -s)\frac{(1-\nu_\mu) r_--r_+(1+\sigma_\mu)}{2 \alpha_\mu}-(\Delta +s-2)\frac{(1-\nu_\mu) r_-+r_+(1+\sigma_\mu)}{2 \alpha_\mu}}
	\end{align}
	For TMG, we have two spin-2 operators corrresponding to the massive and massless modes, for which the conformal dimensions are given by \cite{Li:2008dq,Skenderis:2009nt}
	\begin{align}
		(h,\bar h)=\begin{cases}
			&(2,0)~,~(0,2)~~~~~~~\,\textrm{massless}\\
			&\left(\frac{3}{2}-\frac{\lambda}{2},-\frac{1}{2}-\frac{\lambda}{2}\right)~~\textrm{massive}
		\end{cases}
	\end{align}
	Therefore, for the massive mode, we have the scaling dimension
	\begin{align}
		\Delta^{(m)}=h^{(m)}+\bar h^{(m)}=1-\lambda\,,
	\end{align}
	leading to 
	\begin{align}
		v_B^\textrm{cm}(1-\lambda,2)=\frac{2\pi}{\beta}\frac{\alpha_\mu}{\lambda r_+(1+\sigma_\mu)-r_-(1-\nu_\mu)}
	\end{align}
	Similarly, for the left-moving massless mode, we have
	\begin{align}
		v_B^\textrm{cm}(2,2)=-\frac{2\pi}{\beta}\frac{\alpha_\mu}{r_-(1-\nu_\mu)+r_+(1+\sigma_\mu)}
	\end{align}
	which correctly aligns with \eqref{parameters-comoving}, corroborating our earlier holographic analyses.
\section{Rotating shockwave: fast scrambling}\label{Sec.Rotating shockwave}

As pointed out in \cite{Malvimat:2021itk}, the shockwave with zero angular momentum in fact cannot escape from the asymptotic boundary and fall into the black hole's singularity \cite{Cruz:1994ir}. Therefore, the null coordinates \cite{Reynolds:2016pmi} are not suitable for constructing the Dray-'t Hooft solution. This observation inspired \cite{Malvimat:2021itk} to consider shockwaves with non-zero angular momentum. In this section, we conduct a comprehensive investigation of the fast scrambling phenomenon induced by a null shockwave with finite angular momentum, using both the pole-skipping method and the shockwave method, and obtain consistent chaos parameters.

\subsection{Affine coordinates}

First, we construct the affine coordinates corresponding to the rotating shockwave with a non-zero angular momentum at the outer horizon of the deformed black hole following \cite{Malvimat:2021itk}, which helps set up the Dray-'t Hooft solution \cite{Dray:1984ha}. For the sake of completeness, we note down the components of the $\TTbar$ deformed bulk metric \cite{Basu:2025fsf} in the Schwarzschild coordinate system $\left(t,r,x\right)$,
\begin{align}\label{grrgtt}
	&g_{rr}=\frac{r^2}{(r^2-r_+^2)(r^2-r_-^2)}~~,~~g_{tt}=\frac{1}{r_h^2 \alpha_\mu^2}\left( r_-^2(1+\nu_\mu)^2 \left(r^2-r_-^2\right)-r_+^2 (1-\sigma_\mu)^2 \left(r^2-r_+^2\right)\right)\,,\notag\\
	&g_{xx}=\frac{1}{r_h^2 \alpha_\mu^2}\left(r_+^2 (1+\sigma_\mu)^2 \left(r^2-r_-^2\right)-r_-^2 (1-\nu_\mu)^2 \left(r^2-r_+^2\right)\right)\,,\notag\\&
	g_{tx}=\frac{ r_+ r_-}{r_h^2 \alpha_\mu^2}\left(-2 r^2 (\nu_\mu+\sigma_\mu)+(1+\nu_\mu)(1+\sigma_\mu)r_-^2-(1-\nu_\mu)(1-\sigma_\mu)r_+^2 \right)\,.
\end{align}
For the outgoing null rotating geodesic with energy $\mathcal{E}$ and angular momentum $\mathcal{L}$, the tangent vector $\xi^{\mu}$ satisfies\footnote{For the infalling null rotating geodesic, we instead have
	\begin{align}
		\xi^2=0~~,~~\xi\cdot\zeta_E=\CE~~,~~\xi\cdot\zeta_L=-\CL\,.
\end{align}},
\begin{align}
	\xi^2=0~~,~~\xi\cdot\zeta_E=-\CE~~,~~\xi\cdot\zeta_L=\CL\,,
\end{align}
and is given by,
\begin{equation}\label{xi}
	\xi=\frac{\sqrt{g_{tt} \CL^2+\CE \left(2 g_{tx} \CL+\CE g_{xx}\right)}}{\sqrt{g_{rr} \left(g_{tx}^2-g_{tt} g_{xx}\right)}}\del_r+\frac{g_{tx} \CL+\CE g_{xx}}{g_{tx}^2-g_{tt} g_{xx}}\del_t+\frac{g_{tt} \CL+\CE g_{tx}}{g_{tt} g_{xx}-g_{tx}^2}\del_x\,.
\end{equation}
where $\zeta_E=\del_t$ and $\zeta_L=\del_x$ are the Killing vectors associated with time translational symmetry and rotational symmetry. Correspondingly, the co-tangent vector $\xi_\mu$ is given by,
\begin{align}
	\xi^{\pm}_{\mu}\cdot d x^\mu&=\sqrt{\frac{g_{rr} \left(g_{tt} \CL^2+\CE (2g_{tx} \CL+\CE g_{xx})\right)}{g_{tx}^2-g_{tt}g_{xx}}}dr \mp\CE\left( d t-\frac{\CL}{\CE}d x\right),
\end{align}
where the superscript $\pm$ distinguishes the cotangent vectors associated with outgoing and infalling null geodesics, respectively. Then we define two coordinates $\left(u,v\right)$, which are null coordinates at the outer horizon and facilitate the construction of affine coordinates. Their definition is as follows (with $\CE=1$)\footnote{We hope that the same symbol “$\tau$” with different meanings in \eqref{dudv} and Appendix.\ref{Appendix Bounds} will not cause any confusion.},
\begin{equation}\label{dudv}
	u,v=r_\star\mp \tau,\quad  du=\xi^{+}_{\mu}\cdot d x^\mu,\quad dv=\xi^{-}_{\mu}\cdot d x^\mu,
\end{equation}
where $\tau=t-\mathcal{L}\: x$, and $r_\star$ is a radial tortoise coordinate defined as
\begin{equation}\label{torcrot}
	\begin{aligned}
		r_\star=&-\int_{r}^{\infty}dr\sqrt{\frac{g_{rr}\left(g_{tt}\mathcal{L}^2+2g_{t\varphi}\mathcal{L}+g_{\varphi\varphi}\right)}{g_{t\varphi}^2-g_{tt}g_{\varphi\varphi}}}\\
		=&\frac{\Xi_1}{r_h^2}\text{arctanh}\left[\frac{\Xi_1}{\sqrt{h(r)}\alpha_\mu }\right]
		-\frac{\Xi_2}{r_h^2}\text{arctanh}\left[\frac{\Xi_2}{\sqrt{h(r)}\alpha_\mu}\right]\,.
	\end{aligned}
\end{equation}
In the above, the function $h(r)$ is given by
\begin{equation}
	h(r)=g_{tt} \CL^2+2 g_{tx} \CL+g_{xx},
\end{equation}
and we have utilized the shorthand notations,
\begin{equation}
	\Xi_1=r_+\left(1-\sigma_\mu\right)\mathcal{L}-r_-\left(1-\nu_\mu\right),\quad \Xi_2=r_+\left(1+\sigma_\mu\right)-r_-\left(1+\nu_\mu\right)\mathcal{L}\,,
\end{equation} 
to avoid notational clutter.
The above tortoise coordinate $r_\star$ reduce to \eqref{rstar} in the limit of zero angular momentum $\mathcal{L}\rightarrow 0$, as expected. In terms of the coordinates $\left(u,v\right)$, the $\TTbar$ deformed bulk metric \eqref{deformed metric} can be written as, 
\begin{equation}\label{Frline}
	\begin{aligned}
		ds^2
		=&\CF(r)\,d\mathtt{u}d\mathtt{v}+h(r)\left(d x+\bar{h}(r)d\tau\right)^2\\
		=&\CF(r)\,d\mathtt{u}d\mathtt{v}+\hat{h}\left(r\right)\left(d\phi+h_\tau\left(r\right) d\tau\right)^2
	\end{aligned}
\end{equation}
where we introduce the co-moving angular coordinate $\phi$ \eqref{co-moving} in the second line, and $F,\bar{h},\hat{h},h_\tau$ are functions of $r$ given as follows,
\begin{equation}
	\begin{aligned}
		\CF(r)=&\frac{g_{tx}^2-g_{tt}g_{xx}}{g_{tt} \CL^2+2 g_{tx} \CL+ g_{xx}},\quad 
		\bar{h}(r)=\frac{g_{tt} \CL+g_{tx}}{g_{tt}\CL^2+2 g_{tx}\CL+g_{xx}}\\
		\hat{h}\left(r\right)=&\frac{h}{\left(1-\Omega \mathcal{L}\right)^2},\quad h_\tau\left(r\right)=\Omega+\left(1-\Omega \mathcal{L}\right)\bar{h}.
	\end{aligned}
\end{equation}
Note that $h_\tau\left(r\right)$ vanishes at the outer horizon, which implies $u,v$ are indeed null coordinates at the outer horizon, as expected. Furthermore, the tangent vectors $\chi=\del_\mathtt{u}\,,\del_\mathtt{v}$ are not affine, namely,
\begin{equation}\label{geq}
	\chi\cdot\nabla\chi=\CK\chi,\quad \CK=\Big|\frac{1}{2}\xi^{\pm}\cdot\del \CF(r)\Big|=\frac{1}{2}\xi^r\del_r\CF(r).
\end{equation}
The generalized surface gravity $\kappa_s$, which characterizes the blueshift along the ingoing null geodesic and appears in the exponential factor $e^{\kappa_s \tau_0}$ for a particle escaping from the asymptotic boundary at coordinate time $\tau_0$, coincides with the value of $\mathcal{K}$
evaluated at the outer horizon \cite{Malvimat:2021itk}
\begin{equation}\label{Surface-gravity-rotating}
	\kappa_s=\CK\big|_{r=r_+}=\frac{r_h^2}{\Xi_2}=\frac{2\pi}{\beta(1-\Omega\CL)}.
\end{equation}
where we have used \eqref{beta-Omega}.
In the following, we will find that in the $(\tau,\phi)$ coordinates adapted to the the rotating shockwave, the Lyapunov exponent is still given by the surface gravity at the outer horizon, i.e.
\begin{align}\label{rotat Lyapunov exponent}
	\lambda_L=\kappa_s\geq \frac{2\pi}{\beta},
\end{align}
identical to its undeformed, non-anomalous counterpart \cite{Malvimat:2021itk}.
Surprisingly, this Lyapunov exponent is completely independent of the deformation parameter $\mu$ or the Chern-Simons coupling $\lambda$, which seem to indicate that it has nothing to do with the gravitational anomalies or the $\TTbar$ deformation. Interestingly, in order for the rotating shockwave to be released from the asymptotic boundary and fall into the black hole's singularity, the angular momentum cannot be taken arbitrary values \cite{Cruz:1994ir}. Specifically, the bounds of the angular momentum in the $\TTbar$ deformed black hole geometry \eqref{deformed metric}, which depend on the inverse temperature $\beta$ and the angular velocity $\Omega$, are given by \eqref{bounds} in appendix \ref{Appendix Bounds}. 

Now we turn to set up the affine Kruskal-type coordinates at the horizon, which are defined as follows\footnote{Note that the reason we do not insert a negative sign in front of $u$ here is because this $u$ differs by a negative sign from the $u$ in  \eqref{in-out-metric}, and this way the $U$ here remains consistent with the convention in \eqref{UV}.},
\begin{align}
	U=-e^{\kappa_s \mathtt{u}}~~,~~V=e^{\kappa_s \mathtt{v}}\,,\label{Kruskal}
\end{align}
one can easily check that the coordinates $U,V$ are affine at the outer horizon by \eqref{geq} and \eqref{Surface-gravity-rotating}. The right exterior is captured by $U<0<V$, and the left exterior is obtained by reversing the signs of $(U,V)$, i.e. $0<V<U$. In the Kruskal-type coordinate system, the deformed black hole metric \eqref{deformed metric} takes the following form
\begin{align}
	\d s^2=\frac{\CF(UV)}{\kappa_s^2UV}\d U\,\d V+\hat{h}(UV)\left(\d\phi+h_\tau(UV)\frac{U\d V-V\d U}{2\kappa_s U V}\right)^2\,,
\end{align}
However, we can not write down the explicit analytic expressions of $\mathcal{F},\hat{h}$ and $h_\tau$ in terms of $\left(U,V\right)$, since the tortoise coordinate \eqref{torcrot} is too complicated to invert explicitly.

%The coordinates $(\phi,\tau)$ are defined as
%\begin{align}
%    &\phi=x-\Omega t~~,~~\tau=t-\CL x\notag\\
%    \implies &x=\frac{\phi+\Omega\tau}{1-\Omega\CL}~~,~~t=\frac{\tau+\CL\phi}{1-\Omega\CL}\,.\label{co-moving-rotating}
%\end{align} 
%Note that the $\phi$ coordinate is same as the co-rotating coordinate \eqref{co-moving} at the horizon.

\subsection{Shockwave solution: single-particle backreaction}
We now proceed to construct the Dray–'t Hooft shockwave solution, which captures the backreaction of the wormhole geometry induced by the rotating shockwave. Without loss of generality, we mainly focus on the case of a single particle localized on the horizon $U=0$ at $\phi=0$, which corresponds to a perturbation sourced by the following localized stress tensor:
\begin{equation}\label{rotatingshock}
	T_{UU}^{\text{shock}}=E_0e^{\kappa_s \tau_0}\delta\left(U\right)\delta\left(\phi\right),
\end{equation}
where $E_0$ is the energy of the shockwave measured at the asymptotic boundary, and $e^{\kappa_s \tau_0}$ represents the blueshift factor. The Dray-'t Hooft ansatz claimed that this perturbation induces a shift in the coordinate $V$ as
\begin{equation}
	V\rightarrow V+E_0e^{\kappa_s \tau_0} f(\phi)\Theta(U),
\end{equation}
or equivalently,
\begin{align}\label{t-Hooft-ansatz}
	ds^2\rightarrow ds^2+E_0e^{\kappa_s \tau_0}  \frac{\CF\left(UV\right) }{\kappa_s^2 U V}\delta(U)f(\phi)dU^2\,,
\end{align}
Substituting the above perturbation of the metric into \eqref{TMG-EoM}, we can, in principle, calculate the variation of the $UU$ component of the equation of motion of TMG,
\begin{equation}\label{PerEOM}
	\delta E_{UU}+\frac{1}{\lambda}\delta C_{UU}=8\pi G T_{UU}^{\text{shock}}.
\end{equation}
However, as we mention above, the explicit expression of $\mathcal{F}\left(UV\right)$ is not known, which will introduce trouble in our calculations. On the other hand, the left-hand side of \eqref{PerEOM} can be written down explicitly in terms of the radial coordinate $r$, except the part involving $\delta\left(U\right)$. In other words, the left-hand side of \eqref{PerEOM} is given by,
\begin{equation}
	\delta E_{UU}+\frac{1}{\lambda}\delta C_{UU}=\mathcal{G}\left(\mathcal{F},\hat{h},h_\tau\right)\delta\left(U\right)
\end{equation}
where $\mathcal{G}$ is a function of $\mathcal{F},\hat{h}$ and $h_\tau$, and it can be written down implicitly in terms of $r$. For example, the term $\partial_{UV}\mathcal{F}\left(UV\right)$ inside $\mathcal{G}\left(\mathcal{F},\hat{h},h_\tau\right)$ is given by,
\begin{equation}
	\partial_{UV}\mathcal{F}\left(UV\right)=\frac{dr}{d(UV)}\frac{d\mathcal{F}\left(r\right)}{dr}=\left(\frac{d\left(-e^{2\kappa_s r_\star}\right)}{dr}\right)^{-1}\mathcal{F}'\left(r\right)=\left(-2\kappa_s e^{2\kappa_sr_\star}\frac{dr_\star}{dr}\right)^{-1}\mathcal{F}'\left(r\right).
\end{equation} 	
Using this trick, the left-hand side of \eqref{PerEOM} takes the following form in the near horizon limit
\begin{equation}\label{deltaEOMpro}
	\begin{aligned}
		\delta E_{UU}+\frac{1}{\lambda}\delta C_{UU}\propto E_0e^{\kappa_s \tau_0} \delta\left(U\right)\left( f'''\left(\phi\right)-\left(q_++q_-+q_m\right) f''\left(\phi\right)\right.\\
		\left.+\left(q_+ q_-+q_-q_m+q_+q_m\right)f'\left(\phi\right)-q_+q_-q_mf\left(\phi\right)\right)
	\end{aligned}
\end{equation}
where $q_\pm$ and $q_m$ are given by,
\begin{equation}
	\begin{aligned}
		q_{\pm}=r_h\left(\frac{r_-\left(1-\nu_\mu\right)\pm r_+\left(1+\sigma_\mu\right)}{r_h \alpha_\mu}-\frac{r_h\mathcal{L}}{\Xi_2}\right),\\
		q_m=r_h\left(\frac{r_-\left(1-\nu_\mu\right)-\lambda  r_+\left(1+\sigma_\mu\right)}{r_h \alpha_\mu}-\frac{r_h\mathcal{L}}{\Xi_2}\right).
	\end{aligned}
\end{equation}
The proportional factor reads,
\begin{equation}
	\frac{\Xi_2^2\alpha_\mu^3}{r_+^3\left(1+\sigma_\mu\right)^3\chi_1\chi_2}\exp\left(-2\mathcal{M}\text{arctanh}\mathcal{M}\right),
\end{equation}
where
\begin{equation}\label{MCHI}
	\begin{aligned}
		\mathcal{M}=&\Xi_1/\Xi_2,\\
		\chi_1=&r_+\left(1+\mathcal{L}+\sigma_\mu-\mathcal{L}\sigma_\mu\right)-r_-\left(1+\mathcal{L}-\nu_\mu+\mathcal{L}\nu_\mu\right),\\
		\chi_2=&r_+\left(1-\mathcal{L}+\sigma_\mu+\mathcal{L}\sigma_\mu\right)+r_-\left(1-\mathcal{L}-\nu_\mu-\mathcal{L}\nu_\mu\right).
	\end{aligned}
\end{equation}
Substituting \eqref{deltaEOMpro} and \eqref{rotatingshock} into \eqref{PerEOM}, we may now obtain the Dray-'t Hooft solution as follows
\begin{equation}
	f\left(\phi\right)=c_1 e^{q_+\phi}+c_2 e^{q_-\phi}+c_3 e^{q_m \phi },
\end{equation}
where $c_1,c_2$ and $c_3$ are three constants which can be determined by the the falloff boundary condition. According to the form of OTOC \eqref{OTOC-from-h}, the butterfly velocities of the rotating shockwave case may be obtained as follows
\begin{equation}\label{rotat butterfly velocities}
	\begin{aligned}
		\frac{1}{v_B^\pm}=&\frac{\Xi_2}{r_h}\left(\frac{r_-\left(1-\nu_\mu\right)\pm r_+\left(1+\sigma_\mu\right)}{r_h \alpha_\mu}-\frac{r_h\mathcal{L}}{\Xi_2}\right),\\ \frac{1}{v_B^m}=&\frac{\Xi_2}{r_h}\left(\frac{r_-\left(1-\nu_\mu\right)-\lambda  r_+\left(1+\sigma_\mu\right)}{r_h \alpha_\mu}-\frac{r_h\mathcal{L}}{\Xi_2}\right).
	\end{aligned}
\end{equation}
We refrain from reporting the expressions for the butterfly velocities in terms of $\beta$ and $\Omega$ since they are cumbersome and not quite illuminsting. However, one can easily check the above results will reduce to \eqref{chaos-parameters-Sch-pm} and \eqref{chaos-parameters-Sch-m} in the non rotating limit, i.e. $\mathcal{L}\rightarrow 0$. Furthermore, we can investigate three interesting limits,
\begin{itemize}
	\item[$\bullet$] Undeformed limit, i.e. $\mu\rightarrow 0$:
	\begin{equation}
		v_B^{\pm}=\pm\frac{1\mp\Omega}{1\mp \mathcal{L}},\quad v_B^m=\frac{1-\Omega^2}{\Omega\left(1+\mathcal{L}\lambda\right)-\mathcal{L}-\lambda}.
	\end{equation}
	\item[$\bullet$] Non gravitational anomalous limit, i.e. $\lambda \rightarrow \infty$:
	\begin{equation}
		v_B^{\pm}=-\frac{\left(1-\Omega^2\right)\left(8\pi^2\mu+\beta_+\beta_-\left(\sqrt{1-\frac{8\pi^2\mu}{\beta_+\beta_-}}-1\right)\right)}{8\pi^2\mu\left(\mathcal{L}-\Omega\right)\pm \beta_+\beta_-\left(1\mp\Omega\right)\left(1\pm \mathcal{L}\right)\left(\sqrt{1-\frac{8\pi^2\mu}{\beta_+\beta_-}}-1\right)},\:
		v_B^m=0.
	\end{equation}
	indicating the loss of the massive mode.
	\item[$\bullet$] Undeformed and non gravitational anomalous limit i.e. $\mu\rightarrow 0$ and $\lambda \rightarrow \infty$,
	\begin{equation}
		v_B^{\pm}=\pm\frac{1\mp\Omega}{1\mp \mathcal{L}},\quad v_B^m=0.
	\end{equation}
\end{itemize}

\subsection{Pole skipping}

In this subsection, we investigate the chaos parameters for the rotating shockwave via the pole-skipping method following Sec.\ref{Poleskipping}. For the pole-skipping analysis, it is customary to work in the infalling Eddington-Finkelstein coordinate system $\left(v,r,\phi\right)$, in which the metric reads,
\begin{align}
	ds^2=&\frac{r^2\Xi_1^2}{\left(r^2-r_-^2\right)^2\Xi_2^2}dr^2-\frac{r_h^2\left(r^2-r_+^2\right)}{\Xi_2^2}dv^2\notag\\
	&+\frac{r_+^2\left(1+\sigma_\mu\right)^2\left(r^2\chi_1\chi_2+\chi_3\chi_4\right)}{r_h^2 \alpha_\mu^2\Xi_2^2}d\phi^2+\frac{2r r_h\sqrt{r^2\chi_1\chi_2+\chi_3\chi_4}}{\left(r^2-r_-^2\right)\Xi_2^2}dvdr\\
	&-\frac{2r_+r_h\left(r^2-r_+^2\right)\left(1+\sigma_\mu\right)\Xi_1}{r_h \alpha_\mu\Xi_2^2}dvd\phi+\frac{2rr_+\Xi_1\left(1+\sigma_\mu\right)\sqrt{r^2\chi_1\chi_2+\chi_3\chi_4}}{r_h \alpha_\mu \left(r^2-r_-^2\right)\Xi_2^2}drd\phi,\notag
\end{align}
where $\chi_3$ and $\chi_4$ are given by,
\begin{equation}
	\begin{aligned}
		\chi_3=&\:r_+^2\mathcal{L}\left(1-\sigma_\mu\right)-r_-^2\mathcal{L}\left(1+\nu_\mu\right)+r_+r_-\left(\nu_\mu+\sigma_\mu\right),\\
		\chi_4=&\:r_+^2\mathcal{L}\left(1-\sigma_\mu\right)+r_-^2\mathcal{L}\left(1+\nu_\mu\right)-r_+r_-\left(2+\sigma_\mu-\nu_\mu\right).
	\end{aligned}
\end{equation}
Note the unusual feature of the above metric, namely $g_{rr}\neq 0$.

As earlier, we consider the following ansatz to describe the perturbation of the background metric on the longitudinal modes or the sound modes,
\begin{equation}\label{metricper}
	g_{\mu\nu}\rightarrow g_{\mu\nu}+h_{\mu\nu}=g_{\mu\nu}+\delta g_{\mu\nu}\left(r\right)e^{-i\left(\omega v-k \phi\right)}.
\end{equation}
We may again choose the radial gauge $\delta g_{r\mu}=0$ for the metric perturbations. Since the chaos parameters are only sensitive to the near-horizon behavior, we can expand the modes $\delta g_{\mu\nu}\left(r\right)$ around the outer horizon as,
\begin{equation}
	\delta g_{\mu\nu}\left(r\right)=\delta g_{\mu\nu}^{\left(0\right)}+\left(r-r_+\right)\delta g_{\mu\nu}^{\left(1\right)}+\mathcal{O}\left(r-r_+\right),
\end{equation}
Subsequently, the equations of motion \eqref{TMG-EoM} for the perturbations \eqref{metricper} can be expanded around the outer horizon as follows,
\begin{equation}\label{EOMrotat}
	\delta \text{EOM}_{vv}=e^{-i\left(\omega v-k \phi\right)}\left(c_{vv}^{\left(0\right)}\delta g^{\left(0\right)}_{vv}+c_{v\phi}^{\left(0\right)}\delta g^{\left(0\right)}_{v\phi}+c_{\phi\phi}^{\left(0\right)}\delta g^{\left(0\right)}_{\phi\phi}+c_{vv}^{\left(1\right)}\delta g^{\left(1\right)}_{vv}+c_{v\phi}^{\left(1\right)}\delta g^{\left(1\right)}_{v\phi}+c_{\phi\phi}^{\left(1\right)}\delta g^{\left(1\right)}_{\phi\phi}\right),
\end{equation}
where the coefficients are listed in \eqref{coeffrotat}. The pole-skipping points can now be obtained by requiring all the coefficients to vanish, i.e.
\begin{equation}
	\begin{aligned}
		\omega_\star^{\text{rot}}=&\frac{i r_h^2}{\Xi_2},\\
		k_\star^{\text{rot}}=&i r_h\left(\frac{r_-\left(1-\nu_\mu\right)\pm r_+\left(1+\sigma_\mu\right)}{r_h \alpha_\mu}-\frac{r_h\mathcal{L}}{\Xi_2}\right)
		,\\
		&i r_h\left(\frac{r_-\left(1-\nu_\mu\right)-\lambda r_+\left(1+\sigma_\mu\right)}{r_h \alpha_\mu}-\frac{r_h\mathcal{L}}{\Xi_2}\right).
	\end{aligned}
\end{equation}
utilizing \eqref{Pole-skipping-point}, one can easily verify that the above results agree with those obtained via the shcokwave method, i.e. \eqref{rotat Lyapunov exponent} and \eqref{rotat butterfly velocities}. Furthermore, from the relations between the coordinates $\left(\tau,\phi\right)$ associated with the rotating shockwave and the Schwarzschild coordinates $\left(t,x\right)$,
\begin{equation}\label{co-moving-rotating}
	\tau=t-\mathcal{L}\: x,\quad \phi=x-\Omega \:t,
\end{equation}
the pole-skipping frequencies in the two coordinate systems are related as follows,
\begin{equation}
	\begin{aligned}
		\omega_{\star}^{\text{Sch}}=\omega_{\star}^{\text{rot}}+\Omega \: k_{\star}^{\text{rot}}=&\pm \frac{i \left(r_-\left(1+\nu_\mu\right)\pm r_+\left(1-\sigma_\mu\right)\right)}{\alpha_\mu},\quad - \frac{i \left(\lambda r_-\left(1+\nu_\mu\right)- r_+\left(1-\sigma_\mu\right)\right)}{\alpha_\mu},\\
		k_\star^{\text{Sch}}=k_\star^{\text{rot}}+\mathcal{L}\:\omega_\star^{\text{rot}}=&\frac{i \left(r_-\left(1-\nu_\mu\right)\pm r_+\left(1+\sigma_\mu\right)\right)}{\alpha_\mu},\quad \frac{i \left(r_-\left(1-\nu_\mu\right)-\lambda  r_+\left(1+\sigma_\mu\right)\right)}{\alpha_\mu}.
	\end{aligned}
\end{equation}
Substituting the above expressions into \eqref{Pole-skipping-point}, we obtain the chaos parameters corresponding to the rotating shockwave in the Schwarzschild coordinate system,
\begin{equation}
	\begin{aligned}
		\lambda_L^{\text{rot},\pm}=&\pm \frac{ \left(r_-\left(1+\nu_\mu\right)\pm r_+\left(1-\sigma_\mu\right)\right)}{\alpha_\mu},\quad v_B^{\text{rot},\pm}=\pm  \frac{r_-\left(1+\nu_\mu\right)\pm r_+\left(1-\sigma_\mu\right)}{r_-\left(1-\nu_\mu\right)\pm r_+\left(1+\sigma_\mu\right)},\\
		\lambda_L^{\text{rot},m}=&	- \frac{ \left(\lambda r_-\left(1+\nu_\mu\right)- r_+\left(1-\sigma_\mu\right)\right)}{\alpha_\mu},\quad v_B^{\text{rot},m}=-\frac{\lambda r_-\left(1+\nu_\mu\right)-r_+\left(1-\sigma_\mu\right)}{r_-\left(1-\nu_\mu\right)-\lambda r_+\left(1+\sigma_\mu\right)}.
	\end{aligned}
\end{equation}
Interestingly, when expressing the above results in terms of $\left(\beta,\Omega\right)$, one can easily confirm that they exactly reproduce \eqref{chaos-parameters-Sch-pm} and \eqref{chaos-parameters-Sch-m}. This implies that fast scrambling may be merely a coordinate artifact, which deserves further investigation.

\section{Summary and discussions}\label{sec:summary}

In this work, we investigate the interplay between solvable irrelevant deformations, gravitational anomalies, and quantum chaos in two-dimensional quantum field theories and their holographic duals. We focus on $\TTbar$-deformed anomalous CFT$_2$s and their realization in topologically massive gravity (TMG), where parity violation induces an intrinsic left-right asymmetry in the dynamics.

%We began by reviewing the structure of the $\TTbar$ deformation in the presence of gravitational anomalies and its holographic implementation via mixed boundary conditions in TMG. This framework provides a controlled setting in which both the deformation and the anomaly can be treated nonperturbatively, leading to a deformed BTZ black hole geometry that captures the thermodynamic properties of the boundary theory. The deformation modifies the energy spectrum and introduces a finite scale $\mu$, while the anomaly affects the relation between left- and right-moving sectors.

We analyzed the chaotic properties of the deformed theory using two complementary approaches: pole-skipping and shock-wave geometries. From the pole-skipping analysis of the retarded Green’s function of the energy-momentum tensor, we identified special points in complex frequency-momentum space that encode the Lyapunov exponent and butterfly velocity. Our results show that, despite the presence of both irrelevant deformation and gravitational anomaly, the (average) Lyapunov exponent still saturates the universal chaos bound with a compact spatial coordinate. At the same time, the butterfly velocity exhibits nontrivial dependence on both the $\TTbar$ deformation parameter and the anomaly. In particular, the chiral nature of the theory leads to an asymmetry in the propagation of perturbations, which is reflected in the structure of the pole-skipping points as well as in the shock-wave analysis. This provides a clear illustration of how parity-violating dynamics can modify the spatial spread of chaos without altering its temporal growth rate.

To further elucidate these effects, we constructed the rotating shock-wave geometry in the rotating deformed BTZ background, following \cite{Malvimat:2021itk}. We found that the deformation modifies the relation between boundary time and Kruskal coordinates, as well as the effective coupling controlling the backreaction of infalling perturbations. Nevertheless, the essential mechanism of fast scrambling remains intact, and the exponential sensitivity to initial conditions persists in the deformed geometry. Incidentally, the upper bound to the shockwave's angular momentum $\CL$ corresponds exactly to the (left-moving) butterfly speed in the Schwarzschild frame (cf. appendix \ref{Appendix Bounds}), leading to the following stricter bound for the effective Lyapunov exponent in the frame adapted to the null rotating shockwave
\begin{align}
	\lambda_L\leq \frac{2\pi}{\beta\left(1-\frac{\Omega}{v_B^\textrm{+}}\right)}\,.
\end{align}
At face value, this looks surprisingly similar to the bound obtained in \cite{Halder:2019ric}, where the author conjectured the following bound
\begin{align}
	\lambda_L\leq\frac{2\pi}{\beta\left(1-\left|\Theta/\Theta_c\right|\right)}
\end{align}
for a thermodynamic ensemble at inverse temperature $\beta$ and chemical potential $\Theta<\Theta_c$ corresponding to a global symmetry. However, this analogy with a critical angular chemical potential should not be overinterpreted: $v_B$ is a dynamical scrambling velocity, not a critical value of $\Omega$, and the resemblance is likely accidental.

Taken together, these results highlight the robustness of holographic chaos under deformations of both the ultraviolet and infrared structure of the theory. The $\TTbar$ deformation, despite being irrelevant, preserves the universal features of chaotic dynamics, while the gravitational anomaly introduces controlled and physically meaningful modifications. From a broader perspective, this suggests that the near-horizon origin of chaos is insensitive to many details of the UV completion, provided the deformation can be implemented consistently within the holographic framework.

There are several interesting directions for future work. One natural extension is to explore more general classes of irrelevant deformations, such as $J\bar T$ or higher-spin analogues, with  gravitational anomalies included, and to understand how they affect chaotic observables. It would also be interesting to investigate the role of warped AdS$_3$ geometries in TMG, where the dual field theories are expected to be warped conformal field theories rather than standard CFTs. 

An important future direction lies in the behavior of mutual information scrambling in the deformed anomalous geometry. In ordinary Einstein gravity, the scrambling of mutual information across rotating BTZ wormholes may be analyzed using HRT surfaces \cite{Hubeny:2007xt} in the shockwave geometries \cite{Roberts:2014ifa,Shenker:2013pqa,Malvimat:2021itk}. However, in the present setup with gravitational anomaly, the holographic entanglement functional receives an additional contribution from the twist of the normal frame along the bulk worldline, which can be equivalently described by the dynamics of a spinning particle \cite{Castro:2014tta}. Consequently, computing of mutual information for wormhole-crossing extremal surfaces requires a proper treatment of the Lorentzian transport of this frame twist throughout the shockwave geometry. At present, no general technique is available for systematically evaluating this anomalous contribution from the twist along the worldline in such time-dependent wormhole-crossing backgrounds, and we leave this problem for future investigation.
%Another promising direction is to study subleading corrections in the large-$N$ expansion, which may reveal how quantum effects modify the pole-skipping structure and the shock-wave analysis.

\section*{Acknowledgment}
We would like to thank Prof. Qiang Wen for useful discussions and comments on the draft. MX is supported by SEU Innovation Capability Enhancement Plan for Doctoral Students     (Grant No.CXJH$\_$SEU 26163). We also acknowledge the use of Tom Hartman's excellent Mathematica package ``GREATER2".

\appendix	

\section{Bounds on shockwave angular momentum in $\TTbar$ deformed spacetime}\label{Appendix Bounds}

In this appendix, we investigate the range of angular momentum values of the null geodesics that can release from the asymptotic boundary and fall into the black hole's singularity in the $\TTbar$ deformed spacetime \eqref{deformed metric}. The Killing vectors $\tilde{\zeta}_E=\partial_\tau$ and $\tilde{\zeta}_L=\partial_\varphi$ of the auxiliary BTZ black hole geometry \eqref{BTZgeometry} can be expanded in terms of those of $\TTbar$ deformed spacetime \eqref{deformed metric}, i.e. $\zeta_E=\partial_t$ and $\zeta_L=\partial_x$, by the dynamical coordinate transformations \eqref{coordinatetrans},
\begin{equation}\label{expansion}
	\begin{aligned}
		\tilde{\zeta}_E=&\:\left(1+\mu\left(\kappa \mathcal{L}_\mu+\bar{\kappa}\bar{\mathcal{L}}_\mu\right)\right)\zeta_E-\mu\left(\kappa\mathcal{L}_\mu-\bar{\kappa}\bar{\mathcal{L}}_\mu\right)\zeta_{L},\\
		\tilde{\zeta}_L=&\:\mu\left(\kappa\mathcal{L}_\mu-\bar{\kappa}\bar{\mathcal{L}}_\mu\right)\zeta_E+ \left(1-\mu\left(\kappa \mathcal{L}_\mu+\bar{\kappa}\bar{\mathcal{L}}_\mu\right)\right)\zeta_{L}.
	\end{aligned}
\end{equation}
For null geodesics releasing from the asymptotic boundary and falling into the black hole singularity, the angular momentum in the auxiliary BTZ geometry \eqref{BTZgeometry} must satisfy the constraint
\begin{equation}\label{conditionfallinto}
	\frac{2 r_-r_+}{r_+^2+r_-^2}\leq\frac{\tilde{\mathcal{L}}}{\tilde{\mathcal{E}}}\leq 1,
\end{equation}
where $\tilde{\mathcal{E}}$ and $\tilde{\mathcal{L}}$ are the energy and the angular momentum of the null geodesics, respectively, i.e.,
\begin{equation}\label{ELA}
	\tilde{\mathcal{E}}=-\xi\cdot \tilde{\zeta}_E,\quad \tilde{\mathcal{L}}=\xi\cdot \tilde{\zeta}_L,
\end{equation}
where $\xi$ is the tangent vector of the null geodesics. Since we do not introduce the radial coordinate transformation for the $\TTbar$ deformation in the bulk, we can obtain the bounds of the angular momentum $\mathcal{L}$ by expressing $\tilde{\mathcal{E}}$ and $\tilde{\mathcal{L}}$ in terms of $\mathcal{E}$ and $\mathcal{L}$. More explicitly, substituting \eqref{expansion} and \eqref{ELA} into \eqref{conditionfallinto}, we arrive at,
\begin{equation}
	\frac{2r_- r_+}{r_+^2+r_-^2}\leq \frac{\left(r_+^2\left(1-\sigma_\mu\right)-r_-^2\left(1+\nu_\mu\right)\right)\mathcal{L}+r_+r_-\left(\nu_\mu+\sigma_\mu\right)\mathcal{E}}{-r_+r_-\left(\nu_\mu+\sigma_\mu\right)\mathcal{L}+\left(r_+^2\left(1+\sigma_\mu\right)-r_-^2\left(1-\nu_\mu\right)\right)\mathcal{E}}\leq 1,
\end{equation}
or equivalently,
\begin{equation}\label{bounds}
	\frac{r_+r_-\left(2-\nu_\mu+\sigma_\mu\right)}{r_+^2\left(1-\sigma_\mu\right)+r_-^2\left(1+\nu_\mu\right)}\leq \frac{\mathcal{L}}{\mathcal{E}}\leq \frac{r_+\left(1+\sigma_\mu\right)+r_-\left(1-\nu_\mu\right)}{r_+\left(1-\sigma_\mu\right)+r_-\left(1+\nu_\mu\right)},
\end{equation}
where we have adopted the notations \eqref{sigma-nu}. One can also express the above bounds in terms of $\left(\beta,\Omega\right)$ by \eqref{r-pm}. For example, the upper bound is given by,
\begin{equation}
	\frac{1+\frac{4\pi^2 \mu}{\beta_+\beta_- \lambda}+\Omega\sqrt{1-\frac{8\pi^2 \mu}{\beta_+\beta_-}+\left(\frac{4\pi^2 \mu}{\beta_+\beta_- \lambda}\right)^2}}{\Omega\left(1+\frac{4\pi^2 \mu}{\beta_+\beta_- \lambda}\right)+\sqrt{1-\frac{8\pi^2 \mu}{\beta_+\beta_-}+\left(\frac{4\pi^2 \mu}{\beta_+\beta_- \lambda}\right)^2}}.
\end{equation}
Remarkably, this quantity coincides precisely with the inverse of the butterfly velocity \eqref{chaos-parameters-Sch-pm} of the left-moving sector in Schwarzschild coordinates.

\section{Coefficients in \eqref{linearized-Cotton} and \eqref{EOMrotat}:}\label{AppA}
In the following, we list the coefficients appearing in the near horizon expansion of the linearized Cotton tensor \eqref{linearized-Cotton}:
%\begin{align}
%	c_{\mathbf{v}\mathbf{v}}^{(0)}=&i \alpha_\mu^3 k^3 r_h^4+\alpha_\mu^2 k^2 r_-(1-\nu_\mu) r_h^2\left(2 r_h^2-i r_+ (1+\sigma_\mu) \omega \right)\notag\\&+\alpha_\mu  k r_h^2 r_+ (1+\sigma_\mu) \left(\omega  \left(2 r_+^2 (1+\sigma_\mu)^2-3 (1-\nu_\mu)^2 r_-^2\right)-i r_h^2 r_+ (1+\sigma_\mu)\right)\notag\\&-r_-(1-\nu_\mu)r_+^2 (1+\sigma_\mu)^2 \omega  \left(-(1-\nu_\mu)^2 r_-^2 \omega +i r_h^2 r_+ (1+\sigma_\mu)\right)\notag\\
%	c_{\mathbf{v}\phi}^{(0)}=&\frac{2 \alpha_\mu r_h^3 \left(r_h^2+i r_+ (1+\sigma_\mu) \omega \right) \left(\alpha_\mu^2 k^2 r_h^2+\alpha  k r_- (\nu_\mu -1)r_+ (1+\sigma_\mu) \omega -i r_+^3 (1+\sigma_\mu)^3 \omega \right)}{r_+ (1+\sigma_\mu)}\notag\\
%	c_{\phi\phi}^{(0)}=&\frac{\alpha ^2 r_h^3 \omega  \left(r_h^2+i r_+ (1+\sigma_\mu) \omega \right) \left(\alpha  k r_h^2-r_-(1-\nu_\mu)  r_+ (1+\sigma_\mu) \omega \right)}{r_+ (1+\sigma_\mu)}\notag\\
%	c_{\mathbf{v}\mathbf{v}}^{(1)}=&-\alpha_\mu k 
%	^5 r_+ (1+\sigma_\mu)^2 \left(r_+ (1+\sigma_\mu) \omega -i r_h^2\right)\notag\\
%	c_{\mathbf{v}\phi}^{(1)}=&-\alpha_\mu r_h^5 r_+ (1+\sigma_\mu)^2 \omega  \left(r_+ (1+\sigma_\mu) \omega -i r_h^2\right)\notag\\
%	c_{\phi\phi}^{(1)}=&0\,.
%\end{align}
%\textcolor{red}{corrected:}
	\begin{equation}
		\begin{aligned}
			c_{vv}^{\left(0\right)}=&-r_+\left(1+\sigma_\mu\right)\left(ir_h^4 k^3 \alpha^3_\mu-\omega r_- r_+^2\left(1-\nu_\mu\right)\left(1+\sigma_\mu\right)^2\left(i r_+ r_h^2\left(1+\sigma_\mu\right)-\omega r_-^2\left(1-\nu_\mu\right)^2\right)\right.\\
			&\left.-ir_-k^2r_h^2 \alpha_\mu^2\left(1-\nu_\mu\right)\left(2ir_h^2+\omega r_+\left(1+\sigma_\mu\right)\right)\right.\\
			&\left.+kr_+r_h^2\alpha_\mu\left(1+\sigma_\mu\right)\left(-i r_h^2 r_+\left(1+\sigma_\mu\right)-3\omega r_-^2\left(1-\nu_\mu\right)^2+2\omega r_+^2\left(1+\sigma_\mu\right)^2\right)\right),\\
			c_{v\phi}^{\left(0\right)}=&-2\alpha_\mu r_h^2\left(r_h^2+i \omega r_+\left(1+\sigma_\mu\right)\right)\left(k^2\alpha_\mu^2 r_h^2-k r_+r_-\alpha_\mu \omega \left(1-\nu_\mu\right)\left(1+\sigma_\mu\right)-i\omega r_+^3\left(1+\sigma_\mu\right)^3\right),\\
			c_{\phi\phi}^{\left(0\right)}=&-i\omega r_h^2\alpha_\mu^2\left(\omega r_+\left(1+\sigma_\mu\right)-i r_h^2\right)\left(kr_h^2 \alpha_\mu-\omega r_-r_+\left(1-\nu_\mu\right)\left(1+\sigma_\mu\right)\right),\\
			c_{vv}^{\left(1\right)}=&k \alpha_\mu  r_+^2r_h^4\left(1+\sigma_\mu\right)^3\left(\omega r_+\left(1+\sigma_\mu\right)-i r_h^2\right),\\
			c_{v\phi}^{\left(1\right)}=&\omega \alpha_\mu r_+^2r_h^4\left(1+\sigma_\mu\right)^3\left(\omega r_+\left(1+\sigma_\mu\right)-i r_h^2\right),\\
			c_{\phi\phi}^{\left(1\right)}=&0.
		\end{aligned}
	\end{equation}
%The locations of the inner and outer horizons may be expressed in terms of the temperature and angular potential:
%\begin{align}
%	r_+&=\frac{2 \pi  (1+\lambda  \,\Omega)}{\beta  \left(1-\lambda ^2\right) \left(1-\Omega ^2\right)}-\frac{\beta  \lambda  (\lambda +\Omega )}{2 \pi  \left(1-\lambda ^2\right) \mu}\left(1-\sqrt{1-\frac{8 \pi ^2 \mu }{\beta _+ \beta _-}+\left(\frac{4\pi^2\mu}{\beta_+\beta_-\lambda}\right)^2}\right)\,,\notag\\
%	r_-&=\frac{2 \pi  (\lambda +\Omega )}{\beta  \left(1-\lambda ^2\right) \left(1-\Omega ^2\right)}-\frac{\beta  \lambda  (1+\lambda \,\Omega)}{2 \pi  \left(1-\lambda ^2\right) \mu}\left(1-\sqrt{1-\frac{8 \pi ^2 \mu }{\beta _+ \beta _-}+\left(\frac{4\pi^2\mu}{\beta_+\beta_-\lambda}\right)^2}\right)\,.
%\end{align}
%We may trade the dependence on $r_\pm$ with those of $(\beta,\Omega)$ in the above expressions.

The coefficients arising in the near horizon expansion of the Einstein equations \eqref{EOMrotat} are listed below:

\begin{equation}\label{coeffrotat}
	\begin{aligned}
		c_{vv}^{\left(0\right)}=&\frac{k\alpha_\mu\left(kr_h\alpha_\mu\Xi_2+2ir_+r_h\Xi_1\left(1+\sigma_\mu\right)\right)-ir_+^2\left(1+\sigma_\mu\right)^2\chi_1\chi_2\omega}{2r_+^2r_h\Xi_2\left(1+\sigma_\mu\right)^2}\\
		&-\frac{i}{4r_+^3r_h^4\lambda \Xi_2^2\left(1+\sigma_\mu\right)^3}\left(2r_h^4 k^3\alpha_\mu^3\Xi_2^2\right.\\
		&\left.+2r_+^3r_h^2\Xi_1\left(1+\sigma_\mu\right)^3\chi_1\chi_2\omega+2r_+k^2r_h^2 \alpha_\mu^2\Xi_1\Xi_2\left(1+\sigma_\mu\right)\left(2i r_h^2+\Xi_2 \omega\right)\right.\\
		&\left.-r_h^2 r_+^2 k\alpha_\mu\left(1+\sigma_\mu\right)^2\left(r_h^2\left(\chi_1^2+\chi_2^2\right)+4i\Xi_2\chi_1\chi_2\omega\right)\right),\\
		c_{v\phi}^{\left(0\right)}=&\frac{\alpha_\mu\left(\Xi_2\omega-ir_h^2\right)\left(kr_h^2 \alpha_\mu\left(-ikr_h\alpha_\mu +r_+r_h\lambda\left(1+\sigma_\mu\right)\right)-r_+\left(1+\sigma_\mu\right)\left(ikr_h\alpha_\mu\Xi_1+r_+r_h\Xi_2\left(1+\sigma_\mu\right)\right)\omega\right)}{r_+^3 r_h^3\lambda \Xi_2\left(1+\sigma_\mu\right)^3},\\
		c_{\phi\phi}^{\left(0\right)}=&\frac{\alpha_\mu^2\omega \left(-r_h^2-i\Xi_2 \omega\right)\left(k r_h^2 \alpha_\mu+r_+\left(1+\sigma_\mu\right)\left(i r_h^2 \lambda+\Xi_1 \omega\right)\right)}{2 r_+^3 r_h^2 \lambda \Xi_2 \left(1+\sigma_\mu\right)^3},\\
		c_{vv}^{\left(1\right)}=&\frac{\alpha_\mu k\left(\Xi_2 \omega-i r_h^2\right)}{2 r_+^2  \lambda \left(1+\sigma_\mu\right)},\\
		c_{v\phi}^{\left(1\right)}=&\frac{\alpha_\mu \omega \left(\Xi_2 \omega-i r_h^2\right)}{2 r_+^2  \lambda \left(1+\sigma_\mu\right)},\\
		c_{\phi\phi}^{\left(1\right)}=&0.
	\end{aligned}
\end{equation}

	\bibliographystyle{JHEP}
	\bibliography{TTbarTMG}
\end{document}